\documentclass[useAMS,twocolumn,usenatbib]{mnras}
\usepackage{amssymb}
\usepackage{amsmath}
\usepackage{grffile}
\usepackage{graphicx}
\usepackage[british]{babel} 
\usepackage{placeins}
\usepackage{xcolor}
\usepackage{natbib}
\usepackage{appendix}
\usepackage[T1]{fontenc}
\usepackage{ae,aecompl}
\usepackage{lscape}

\def\mnras{MNRAS}

\def\aap{A\&A}

\def\jcap{Journal of Cosmology and Astroparticle Physics}
\def\physrep{Phys. Rept.}

\def\beq{\begin{equation}}
\def\eeq{\end{equation}}
\def\bey{\begin{eqnarray}}
\def\eey{\end{eqnarray}}

\def\gpc{\,{\rm Gpc}}
\def\mpc{\,{\rm Mpc}}

\def\dT{\,{$\delta T$}}
\def\nf{\,{$\langle x_{\rm HI} \rangle$}}
\DeclareMathOperator\erfc{erfc}

\title[EoR Signatures on the 21cm 2PCF and 3PCF I]{Signatures of Cosmic Reionization on the 21cm 2- and 3-point Correlation Function I: Quadratic Bias Modeling}

\author[Hoffmann et al.]{
Kai Hoffmann$^{1,2}$\thanks{E-mail: hoffmann@ics.uzh.ch},
Yi Mao$^{1}$\thanks{E-mail: ymao@tsinghua.edu.cn},
Jiachuan Xu$^{1}$, 
Houjun Mo$^{1,3}$, 
Benjamin D. Wandelt$^{4,5,6}$
\\
$^{1}$Department of Astronomy and Tsinghua Center for Astrophysics, Tsinghua University, Beijing 100084, China\\
$^{2}$Institute for Computational Science, University of Zurich, Winterthurerstr. 190, 8057 Z\"urich, Switzerland\\
$^{3}$Department of Astronomy, University of Massachusetts, Amherst, MA 01003-9305, USA\\
$^{4}$Sorbonne Universit\'e, CNRS, UMR 7095, Institut d'Astrophysique de Paris (IAP), 98 bis bd Arago, 75014 Paris, France\\
$^{5}$Sorbonne Universit\'e, Institut  Lagrange  de  Paris  (ILP),  98  bis bd Arago, 75014 Paris, France\ \\
$^{6}$Center for Computational Astrophysics, Flatiron Institute, 162 5th Avenue, New York, NY 10010, USA\\
}

\date{Accepted XXX. Received YYY; in original form ZZZ}
\pubyear{2019}

\begin{document}
\label{firstpage}
\pagerange{\pageref{firstpage}--\pageref{lastpage}}
\maketitle

\begin{abstract}
The three-point correlation function (3PCF) of the 21cm brightness temperature from the Epoch of Reionization
(EoR) probes complementary information to the commonly studied two-point correlation function (2PCF)
about the morphology of ionized regions. We investigate the 21cm 2PCF and 3PCF in configuration space
using semi-numerical simulations and test whether they can be described by the local quadratic bias model.
We find that fits of bias model predictions
for the 2PCF and 3PCF deviate from our measurements by $\sim 20\%$ at scales above the typical
size of ionized regions ($\simeq 30$ Mpc) and at early times with global neutral fractions of
$\langle x_{\rm HI} \rangle \gtrsim 0.7$. 
At later times and smaller scales these deviations increase strongly, indicating a break down
of the bias model. The 2PCF and 3PCF fits of the linear bias parameter agree at the $10\%$ level
for different EoR model configurations. This agreement holds, when adding redshift space distortions
to the simulations. The relation between spatial fluctuations in the matter density and the 21cm signal,
as predicted by the bias model, is consistent with direct measurements of this relation in simulations
for large smoothing scales ($\gtrsim 30$ Mpc). From this latter test we conclude that negative
amplitudes of the 21cm 3PCF result from negative bias parameters, which describe the anti-correlation
between the matter over-densities and the 21cm signal during the EoR. However, a more detailed interpretation
of the bias parameters may require a description of non-local contributions to the bias model.
\end{abstract}



\section{Introduction}
Observations with upcoming radio telescopes, such as the low frequency segment of the Square Kilometre Array
(SKA-Low)\footnote{www.skatelescope.org} and  the Hydrogen Epoch of Reionization Array
(HERA)\footnote{www.reionization.org}, will ring in a new era of cosmology,
allowing us to observe the distribution of neutral hydrogen (HI) in the early universe at unprecedented
scales. These observations will provide novel insights into how the first luminous objects ionized their
surrounding gas. This so-called Epoch of Reionization (EoR) leads to the almost completely ionized universe that we see today and can be characterized by the large-scale HI distribution at different times.
This distribution is traced by the light emitted at the wavelength of 21 centimeter due to the HI hyperfine transition. 
Today we can observe this light as radio signals, since it was redshifted during its journey through the expanding universe \citep{Field58}.
The 21cm signal is expected to show large-scale spatial fluctuations, which reflect the inhomogeneous matter
distribution in the universe. Current EoR models suggest that these fluctuations are overlaid by patches without any 21cm emission,
which are filled with ionized gas around the first luminous objects \citep[e.g.][]{Sokasian04}.
The upcoming 21cm observations contain therefore not only information about how matter density
fluctuations grow with time due to gravity, but also about how the first luminous objects form and
transfer energy to the intergalactic medium \citep[e.g.][]{FOB06, PL12}.
Extracting this information from observations requires a statistical description of the 21cm fluctuations.
Correlation functions have proved to be useful tools for this purpose.
Studies on the two-point correlation function of the 21cm
EoR signal, or the power spectrum as its Fourier space counterpart, revealed its high sensitivity
to the underlying EoR model \citep[e.g.][]{Pober14}. This opens up the possibility to use this
statistics for constraining EoR models with future observations \citep[e.g.][]{Liu16}.
However, a restriction to one or two-point statistics prevents access to information on the morphology of the ionized regions, which is one of the most 
prominent features in simulated maps of the 21cm signal. Only recently studies probe this additional information using the
three-point correlation in Fourier space, i.e. the bispectrum \citep[e.g.][]{Shimabukuro15}. The results revealed a strong
dependence of the bispectrum on the scale, as well as on the shape of the triangles
from which this statistics is measured \citep[][]{Majumdar17}. This additional information
from the bispectrum can strongly improve the constraints on EoR model parameters from power spectrum measurements \citep{Shimabukuro17}.

Deriving such constraints from observations requires predictions for the 21cm statistics,
which can be obtained from either theoretical modeling, or from simulations.
In the latter case a large number of simulations is needed for a sufficiently dense
sampling of the high dimensional EoR parameter space. In addition, these simulations
need to cover a wide dynamical range to accurately predict the formation of ionizing
sources at scales of $< 1$ Mpc and simultaneously the transfer of their radiation to the intergalactic medium on scales of up to $\gtrsim 20$ Mpc. 
Furthermore, the simulation volume needs to be
sufficiently large ($\gtrsim 100$ Mpc) to allow for precise measurements of the chosen statistics.
These requirements can be met currently only with approximate simulation methods or
emulation techniques \citep[e.g.][]{Mesinger11, SchmitPritchard17}.

Alternatively to simulations, predictions for the 21cm correlation functions can be
obtained from theoretical models.
Such models have been presented for the 21cm power spectrum by \citet[][]{FZH04, Lidz07}. 
To our knowledge, however, the only model for the 21cm bispectrum is restricted to the simplistic case
of randomly distributed ionized spheres, which shows similarities with simulation
results only at small scales \citep{Bharadwaj05, Majumdar17}.

The goal of this work is to further develop the theoretical understanding of the two-point 
and, in particular, the three-point correlation functions (hereafter referred to as ``2PCF'' and ``3PCF'', respectively) 
of the 21cm brightness temperature, based on measurements in simulations.
Our investigation of the 21cm 3PCF is performed for the first time in configuration space, which
may simplify the physical interpretation of our results. 
To interpret our measurements, we employ the quadratic bias model, which has been developed
to model the 3PCF of large-scale halo and galaxy distributions \citep{FryGa93}. In the context
of 21cm statistics, this model assumes that the matter density in a given volume of the universe determines the
21cm brightness temperature in that same volume, while the form of this deterministic relation
is characterized by a set of free parameters. This bias model allows for a perturbative
approximation of the 21cm 2PCF and 3PCF, which depend at leading order on the two free bias
parameters $b_1$ and $b_2$.

We test the bias modeling using a set of $200$ realizations of reionization simulations, 
which were performed with the
semi-numerical code 21cmFAST \citep{Mesinger11}. These simulations differ only in their
initial conditions of random density fields, and cover a total volume of $\simeq (4.5 \ \text{Gpc})^3$,
providing robust measurements of the 21cm 2PCF and 3PCF, as well as estimates for their
error covariance. 

This paper is organized as follows.
We start by presenting our simulation in Section \ref{sec:simulation}.
The bias model is introduced in Section \ref{sec:bias_CF}
in which we also verify how well its prediction for the 2PCF and the 3PCF
agree with the measurements in the simulations.
As the second step in the model validation, we compare the
2PCF and 3PCF fits of bias parameters in Section \ref{sec:bias_comparison}.
Using these bias parameters, we further test in that section how well the relation
between the matter and 21 brightness temperature, as predicted by the bias model,
holds using direct measurements in simulations.
We summarize our results and draw conclusions from our analysis in Section \ref{sec:conclusion}.

\section {Simulation}\label{sec:simulation}

Our analysis is based on simulations of three-dimensional maps of the matter density and the
21cm brightness temperature, using the semi-numerical code \texttt{21cmFAST} \citep{Mesinger11}.
For a fast production of 21cm maps, \texttt{21cmFAST} first derives the evolved matter density distribution
using the Zel'dovich approximation, starting from an initial Gaussian random field.
The three-point statistics of such distributions have been shown to deviate significantly
from those derived with full N-body simulations at low redshifts \citep[][]{Scoccimarro97, Leclercq13}.
In order to investigate how strongly our results are affected by the Zel'dovich approximation
we generate additional matter density distributions using the code \texttt{L-PICOLA}. This code employs
the COLA method to approximate N-body simulations. This type of simulations reproduces
two- and three-point statistics of matter distributions from full N-body simulations with percent level
accuracy, while being computationally less expensive \citep{Tassev13, Howlett15}.

The mean matter and baryon densities are set to $\Omega_M = 0.308$ and $\Omega_b =0.048$, 
respectively, with the Hubble parameter $h=0.678$. The scalar spectral index of primordial density
fluctuations is set to $n_s=0.968$ and the initial power spectrum at $z=300$ is obtained from the
\citet{EH98} transfer function. These parameters are identical for the \texttt{L-PICOLA} runs,
besides the initial redshift, which was set to $z=100$. These latter simulations where run down to $z=8.43$
in $80$ time steps.
The ionized regions and the resulting 21cm signals are computed from the matter field in a second step with
a semi-numerical approach, which is based on the excursion set formalism \citep{FZH04}. It uses the argument
that a given region (i.e. simulation cell) is ionized when the fraction of collapsed matter fluctuations
with a mass above $m_{\rm min}$, which reside within a sphere with radius $R < R_{\rm max}$ around the regions center,
passes a certain threshold, i.e.
%
\begin{figure}
\centering\includegraphics[width=11.5 cm, angle=270]{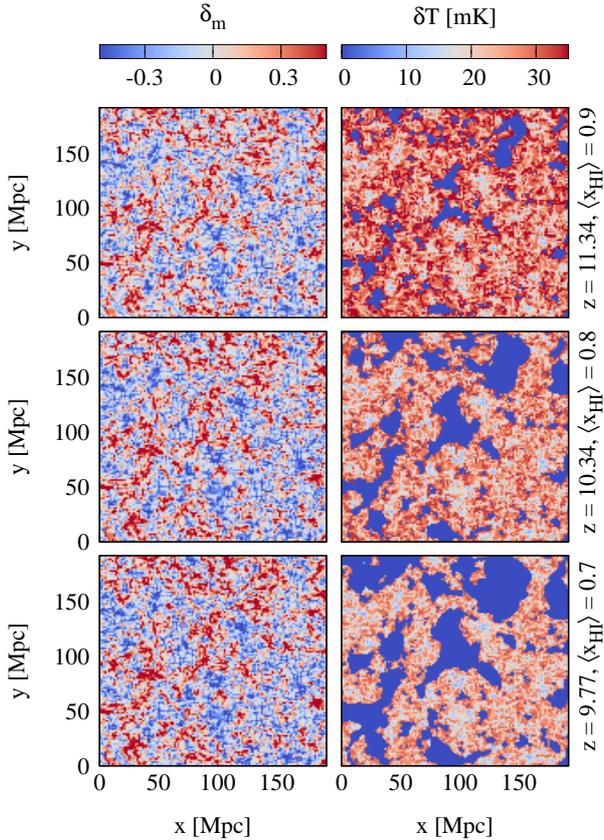}
\caption{
    Simulated dark matter density fluctuations and 21cm brightness temperature from model A without RSD
    (left and right panels respectively) in a $1.5$ \mpc{} slice of a $(200 \mpc)^3$ subvolume in one random realization.
    Results are shown for three redshifts with different global fractions of neutral hydrogen $\langle x_\text{HI} \rangle$.
    Patches of ionized gas without 21cm emission (as represented by blue regions on the right panel) form around
    the highest matter overdensities (as represented by red regions on the left panel) 
    at early times (top panels) and expand with decreasing redshifts (central and bottom panels.)
}
\label{fig:sim}
\end{figure}
%
\begin{equation}
x_{\text{HI}}=\begin{cases}
  1  & \text{if} \ f_{\rm coll} > 1/ \zeta_{\rm ion} \\
  0 & \text{else},
\end{cases}
\label{eq:nf_zeta}
\end{equation}
where $\zeta_{\rm ion}$ is the ionization efficiency parameter and $R_{\rm max}$ can be interpreted as
mean free path length for photons in the intergalactic medium.
The neutral fraction $x_{\text{HI}}$ characterizes the ionization state of that region
and corresponds to its local fraction of HI with respect to its total number
of hydrogen atoms. The collapsed fraction is given by

\begin{equation}
f_{\rm coll} = \erfc
\left(
\frac{\delta_m^{crit} - \delta_m(R)}{\sqrt{2(\sigma_m^2(m_{\rm min}) - \sigma_m^2(R))}}
\right).
    \label{eq:fcoll}
\end{equation}
The normalized spatial fluctuation of the total matter density, $\rho_m$,
around the universal mean $\langle \rho_m  \rangle$ is defined as
\begin{equation}
    \delta_m \equiv (\rho_m - \langle \rho_m \rangle)/ \langle \rho_m \rangle.
	\label{eq:delta}
\end{equation}
Note that the fluctuations $\delta_{\delta T}$, with which we will work later as well, are defined analogously.
We set the critical amplitude of matter fluctuations for spherical collapse to $\delta_m^{crit}=1.68$,
$\delta_m(R)$ is the large-scale matter fluctuation around the potentially ionized region,
smoothed with a Fourier space top-hat filter on the scale $R < R_{\rm max}$ and $\sigma_m^2(R)$ is the variance of
the matter density field for a top-hat filter in configuration space with radius $R$.
The inconsistent filtering for $\delta$ and $\sigma$ allows for faster computations of $\delta_m(R)$
around each simulation cell, since the Fourier space filtering is done by a simple cut of the power spectrum.
A $k$-space top-hat filter is in principle also convenient for theory applications, since for a Gaussian random field,
the differences between $\delta_m(R)$ at different smoothing radii (i.e. the steps of the excursion
set trajectories) are uncorrelated. However, a detailed physical interpretation of the simulation parameters
would require a consistent filtering for $\sigma^2$ and $\delta$ in configuration space, since that is where
the physical processes, which drive the ionization (e.g. the radiation transport) are most conveniently described.
For $\sigma_m^2(m_{\rm min})$ the top-hat filter encloses on average the minimum mass for ionizing fluctuations
$m_{\rm min}$. The minimum mass is derived from the minimum temperature for ionization $T_{\rm min}$ via
\begin{equation}
m_{\rm min} = 7030.97 \ h^{-1}(\Omega_m(z) \ \delta_{nl})^{1/2}
\left(\frac{T_{\rm min}}{\mu(1+z)}\right)^{3/2}
    \label{eq:m_min}
\end{equation}
\citep[e.g.][]{BarkanaLoeb01}, where $\delta_{nl}$ is the non-linear matter fluctuation at virialization,
which was approximated by \citet{BryNo98} as
$\delta_{nl} = 18 \pi^2 + 82 x - 39 x^2$ with $x = \Omega_m(z) - 1$.
The mean molecular weight $\mu$ is set in our simulations to $0.6$
for ionized hydrogen and singly-ionized helium.
Once the neutral fraction has been computed for each simulation cell, the 21cm brightness temperature is obtained via
\begin{equation}
    \delta T = x_{\text{HI}}(1+\delta_m) \left( \frac{T_{\rm s} - T_{\rm CMB}}{T_{\rm s}}  \right) \times C
	\label{eq:dT}
\end{equation}
with
\begin{equation*}
    C =  23.88 \left( \frac{0.15}{\Omega_M \ h^2} \frac{1+z}{10} \right)^{1/2} \\
    \left(\frac{\Omega_b h^2}{ 0.02 }\right) {\rm mK}
\end{equation*}
\citep[e.g.][]{Furlanetto06} for an optically thin inter galactic medium.
The term $x_\text{HI}(1+\delta_m)$ is proportional to the local HI density, assuming a
linear relation between the hydrogen and the full matter densities. \citet{Mesinger11} found that this
assumption breaks down at scales below $1$ Mpc, but holds at larger scales which we study in this work.

We focus on the regime in which the gas has been significantly heated, so that the spin temperature 
$T_{\rm s}$ is much greater than the Cosmic Microwave Background (CMB) temperature $T_{\rm CMB}$. 
This approximation is reasonable after the onset of cosmic reionization when the fluctuations of the 21cm signal are 
dominated by those of the neutral fraction induced by the inhomogeneous reionization
\citep[e.g.][]{ZFH04, Pritchard06, Mesinger11},  but may need to be revisited for a more detailed analysis.

We compute the correlations of the 21cm signal with and without corrections due to redshift space distortions
(hereafter referred to as RSD). Neglecting RSD may be appropriate for analysis of projected 21cm signals in
redshift bins. However, studies on three dimensional 21cm power spectra
found notable contributions on large scales, in particular for early times with high neutral fraction
\citep{McQuinn06, MesingerFurlanetto07},  in which case a detailed modelling of 21cm redshift space distortion
should be taken into account. For selected cases we therefore study the effect of RSD on the 21cm 2PCF and 3PCF,
comparing a simplified RSD prescription implemented in 21cmFAST with a more realistic model MMRRM. 
The MMRRM code \citep{Mao12} is based on the sophisticated investigation of the effects of peculiar velocity on
the observed 21cm signal, which manifests in two aspects: the 21cm brightness temperature is modified, and a
distance separation along line-of-sight is resized in redshift space, both according to the velocity gradient.
The net effect on the 21cm brightness temperature in redshift-space
is $\delta T_b^s ({\bf s}) \propto \left[1+\delta^s_{\rho_{\rm HI}}({\bf s})\right]$, where $\delta^s_{\rho_{\rm HI}}({\bf s})$
is the neutral hydrogen density fluctuations in redshift space. We refer readers to \cite{Mao12} for
a detailed discussion of this methodology. On the other hand, the 21cmFAST code has an option to implement
a simplistic prescription for 21cm RSD, in which a maximum for the absolute value of the velocity gradient,
$|dv/dr|_{max} = 0.2 \ aH$, is enforced by hand to avoid artificial singularity. 

We employ three reionization models which are characterized by different
sets of parameters as summarized in Table \ref{table:EOR_param}.
%
\begin{table}
\begin{center}
\begin{tabular}{ c | c | c }
model & $\zeta$ & $T_{\rm min}$ [K] \\
  \hline			
  A & $60$ & $5\times 10^4$ \\
  B & $60$ & $3\times 10^4$ \\
  C & $40$ & $5\times 10^4$ \\
\end{tabular}
\caption{Sets of reionization model parameters. The fiducial
model A is used in most parts of the analysis, unless
the use of model B and C is explicitly specified.\label{table:EOR_param}}
\end{center}
\end{table}
%
The maximum filtering scale $R_{\rm max}$ is set to $20$ Mpc for all three models.
Model A serves as our fiducial model throughout the analysis unless
the use of model B and C is specified explicitly.

\begin{figure}
\centering\includegraphics[width=6 cm, angle=270]{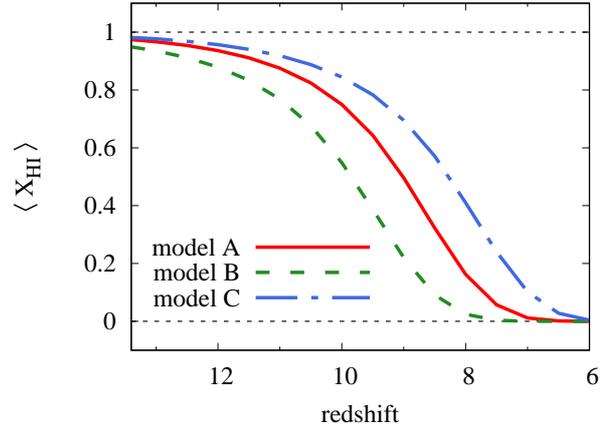}
\caption{
    {Global neutral hydrogen fraction (weighted by volume) versus redshift
    for the three reionization models, defined in Table \ref{table:EOR_param}.
    Results are shown for one realization, while the variation among different
    realizations is at the percentage level.}
    \label{fig:nf-z}
}
\end{figure}

\begin{figure}
\centering\includegraphics[width=7 cm, angle=270]{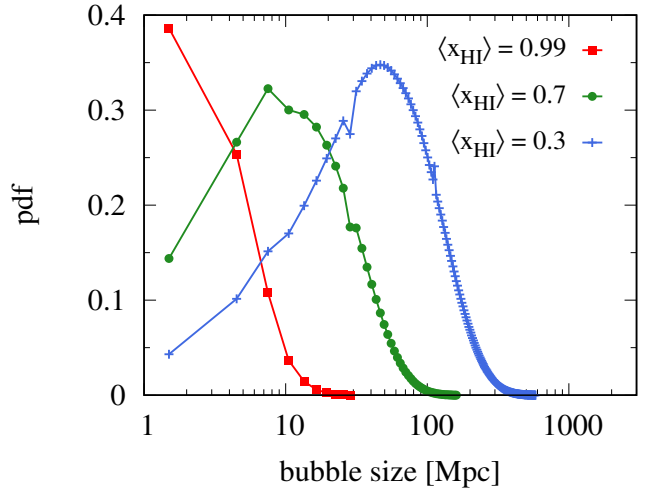}
\caption{Size distribution of ionized regions (referred to as bubbles) from model A,
measured by \texttt{21cmFAST} at three different redshifts
with different global neutral fractions. Dots show mean measurements from different
random realizations. $1 \sigma$ errors are smaller than the symbol size.)
\label{fig:bubble_sizes}}
\end{figure}

All simulations are run in cubical volumes of ($768 \mpc)^3$ with periodic boundary conditions
on Cartesian grids with $1536^3$ ($512^3$) cells and with the side lengths of 
cells of $0.5$ ($1.5$) Mpc, for the matter (HI) fields, respectively.
RSDs are applied along one simulation axis in a plane-parallel approximation.
To estimate the error covariance of our measured statistics, we generate a set of $200$
independent realizations with different random initial conditions for the fiducial model A,
which cover a total volume of $\sim (4.5 \gpc)^3$. For model B and C we
use $100$ realizations to reduce computational costs.
For the \texttt{L-PICOLA} runs we restrict our analysis to just $10$ realizations
with $1536^3$ particles due to the relatively high computational costs compared
to simulation based on the Zel'dovich approximation.
In Fig. \ref{fig:sim} we show the simulated distributions of
the matter density fluctuations in one realization
and the corresponding 21cm brightness temperature distributions from model A without RSDs at different redshifts.
One can see how ionized regions start to form around the largest matter overdensities at high redshifts and
connect with each other, as they grow with decreasing redshift. This growth progresses relatively fast
compared to the growth of the large-scale matter density fluctuations.
The redshift evolution of the global volume-weighted neutral fraction \nf\ is shown in Fig. \ref{fig:nf-z}
for all three reionization models.
This figure illustrates on one hand how
reducing the minimum virial temperature $T_{\rm min}$ of halos that host ionizing sources
in model B from the fiducial value in model A accelerates the global ionization process, as
ionizing sources can now reside in lower mass halos (see equation (\ref{eq:m_min})), which increases
their total number density.
Reducing the reionization efficiency $\zeta$ in model C on the other hand delays the reionization process,
as a larger collapsed fraction is needed for ionizing a given region (see equation (\ref{eq:nf_zeta})).
In Fig. \ref{fig:bubble_sizes}, we show the probability distribution function (pdf) of the sizes of ionized regions
(or HII bubbles), averaged over different random realizations. The sizes are measured by \texttt{21cmFAST} by choosing
random positions in ionized regions and measuring the distance from each of these positions to the
neutral phase in a randomly chosen direction. The distribution of these distances is interpreted
as the bubble size pdf. These pdfs cover scales between $1$ and $200$ Mpc, while their maxima lie
at around $1$, $10$ and $50$ Mpc for \nf{} $=0.99$, $0.7$ and $0.3$, respectively.

\section{Biased correlation functions}\label{sec:bias_CF}

We quantify the clustering of the matter density and 21cm brightness temperature with the 2PCF and 3PCF.
The convenience of these statistics arises from the fact that they are well defined and have been 
extensively investigated in galaxy clustering analysis \citep[e.g.][]{Bernardeau02}. 
The studies therein implied that the correlation functions of some tracers of the total matter density
field can be related to those of the latter via the so-called bias model, as detailed below.

\subsection{Quadratic bias model} \label{sec:bias_model}
In the context of 21cm observations, the local bias model corresponds to a function $F$, which describes a deterministic
relation between the fluctuations in the matter density $\delta_m$ and those in the 21cm brightness temperature $\delta_{\delta T}$,
defined by equation (\ref{eq:delta}), in a region of size $R$ around a given position $\bf x$,
i.e. $\delta_{\delta T}({\bf x}) = F[\delta_m({\bf x})]$.
The term ``local'' refers to the assumption that $\delta_{\delta T}$ is solely determined by
$\delta_m$ in the same region. This assumption can be violated, for instance, by ionizing radiation originating from overdensities 
outside of the region \citep[e.g.][]{FZH04}, or the tidal field induced by the surrounding matter distribution, 
which affects the formation of halos that host the ionizing sources in that region \citep[][]{css12, Baldauf12}.
However, for sufficiently large scales and early stages of reionization, the deterministic
approximation might be adequate, as we will see later in this analysis.

The form of the bias function is unknown at this point, but we can approximate it with a Taylor series
around $\delta_m=0$, assuming that $\delta_{\delta T}$ varies only weakly with $\delta_m$, i.e
$\delta_{\delta T} \simeq \sum_{n=0}^N (b_n/n!)(\delta_m)^n$, where $b_n$ are the bias parameters.
The bias model provides a tool to relate the 2PCF and 3PCF of \dT{} to the corresponding statistics
of the underlying matter field. For establishing these relations for third-order statistics at leading
order in $\delta_m$, we need to expand the bias function at least to second order \citep[][]{FryGa93}, i.e.
\begin{equation}
    \delta_{\delta T} \simeq b_0 + b_1 \delta_m + (b_2/2){\delta_m}^2.
	\label{eq:quad_bias_model}
\end{equation}
Requiring $\langle \delta_{\delta T} \rangle =0$ sets the constrain $b_0=-(b_2/2) {\sigma_m}^2$,
where ${\sigma_m}^2 = \langle {\delta_m}^2 \rangle$ is the variance of the matter density fluctuations. Here
$\langle \ldots \rangle$ denotes the spatial average of the universe. The remaining
free parameters $b_1$ and $b_2$ are referred to as {\it linear} and {\it quadratic} bias parameters, respectively.

\subsection{Two-point correlation functions (2PCF)}\label{sec:2pc}
The 2PCF can be defined via the product of fluctuations at two different positions
\begin{equation}
		\xi(r) \equiv \langle  \delta_1 \delta_2 \rangle(r),
		\label{eq:2pc_def}
\end{equation}
where $\delta_i$ are the fluctuations of some observable at the position ${\bf x}_i$, i.e. $\delta_i \equiv \delta({\bf x}_i)$, and $\langle \ldots \rangle$ denotes the average
over all possible orientations and translations of the $\delta_1\delta_2$ pairs, separated by the distance
$r \equiv |{\bf x}_2-{\bf x}_1|$. We denote the 2PCFs of the matter density fluctuations and 21cm brightness
temperature fluctuations (referred to as ``21cm 2PCF'') as $\xi_m$ and $\xi_{\delta T}$, respectively.
Due to the average over all orientations, the 2PCF is only sensitive to the scale dependence of the clustering,
but not to the shape of large-scale structures. As a second-order statistic, it is also insensitive
to a skewness in the one-point probability distribution of the fluctuations $\delta$.
This additional information can be accessed via the 3PCF, which will be introduced in Section \ref{sec:3pc}.
Our 2PCF measurements are performed on a grid of cubical cells with $6$ Mpc side lengths.
We thereby measure the fluctuations $\delta_m$ and $\delta_{\delta T}$ in each cell and
then search for pairs of cells to obtain the average from equation~(\ref{eq:2pc_def}).

\begin{figure}
\centering\includegraphics[width=11 cm, angle=270]{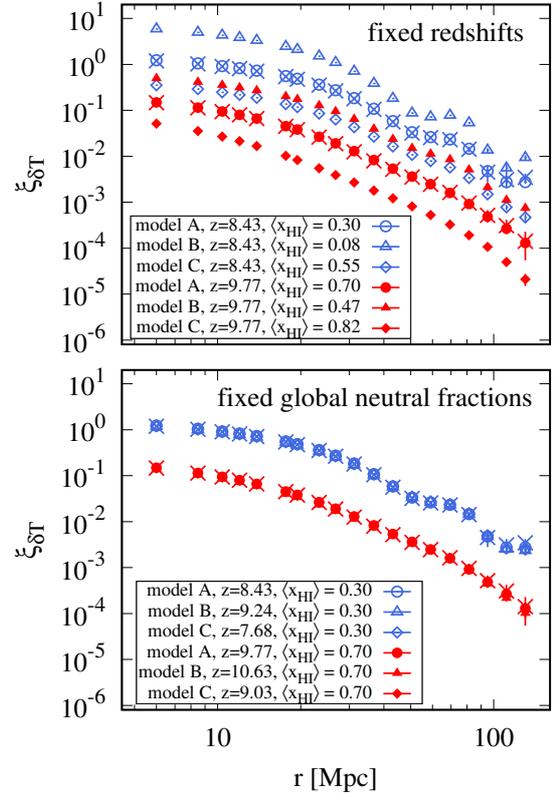}
\caption{{\it Top}: 21cm 2PCF for the three reionization models defined in Table \ref{table:EOR_param},
versus the scale $r$ at two fixed redhifts $z$. The volume weighted global neutral fractions \nf{}
are different for each model and redshift.
Crosses show measurements in simulations with \texttt{L-PICOLA} matter distributions for model A.
Error bars denote $1 \sigma$ uncertainties, which can be smaller than the symbol size.
{\it Bottom}: Corresponding measurements at different redshifts with the same global neutral fractions.
Deviations between measurements from different models are smaller than the symbol size.}
\label{fig:2pc_EORmodels}
\end{figure}

We start our analysis by comparing the mean 21cm 2PCF over all realizations from our three EoR models
without RSD at two fixed redshifts in the top panel of Fig. \ref{fig:2pc_EORmodels}. We find a significant
dependence of the 2PCF on the EoR model at a given redshift as well as a strong redshift evolution
for a given EoR model. Results based on the matter fields from the Zel'dovich approximation and the
\texttt{L-PICOLA} runs (displayed for model A as dots and crosses repsectively) show no significant deviations.

Differences between measurements for different EOR models and redshifts can be approximated
by a single scale independent shift of the amplitude. This behaviour illustrates the limited constraining power
of the 21cm 2PCF on EoR models, which depend on several free parameters. When comparing results at fixed
volume weighted global neutral fractions \nf\ (bottom panel of Fig. \ref{fig:2pc_EORmodels}),
we find only a percent level dependence of the 21cm 2PCF on the EoR model (note that the global neutral
fractions are only fixed at percent level accuracy). This finding
can be understood with the following consideration. The condition for the ionization of a given region
from eq. (\ref{eq:nf_zeta}) can be rewritten with eq. (\ref{eq:fcoll}) in terms of an ionization barrier,
i.e. $\delta_m(R) > B$, with $B \equiv \delta_m^{crit} - D(z) K(\zeta) \sqrt{2}\sqrt{\sigma_{m,0}^2(m_{min}) - \sigma_{m,0}^2(R)}$,
where $K(\zeta) \equiv erf^{-1}(1-\zeta^{-1})$, $\sigma_{m,0}^2$ is the variance of the matter field at $z=0$ and
$D$ is the linear growth factor \citep[e.g.][]{FZH04}. The terms in this expression that depend on $\zeta$ and
$T_{min}$ are degenerate with the redshift dependent growth factor. A certain spatial distribution of neutral fractions $x_{\rm HI}$ can
hence be obtained by different combinations of the redshift and the EoR model parameters. This argument
implies that also $\delta T \propto x_{\rm HI}(1+\delta_m)$ (eq. (\ref{eq:dT})) is roughly model independent
at a fixed global neutral fraction, given that the evolution of $\delta_m$ is very weak at the high redshifts
considered in this work in comparison with the strong evolution of $x_{\rm HI}$ (see Fig. \ref{fig:sim}).
However, this result is specific to the excursion set based simulations, and might break down to some degree
in more complex EoR models.

\begin{figure}
\centering\includegraphics[width=11 cm, angle=270]{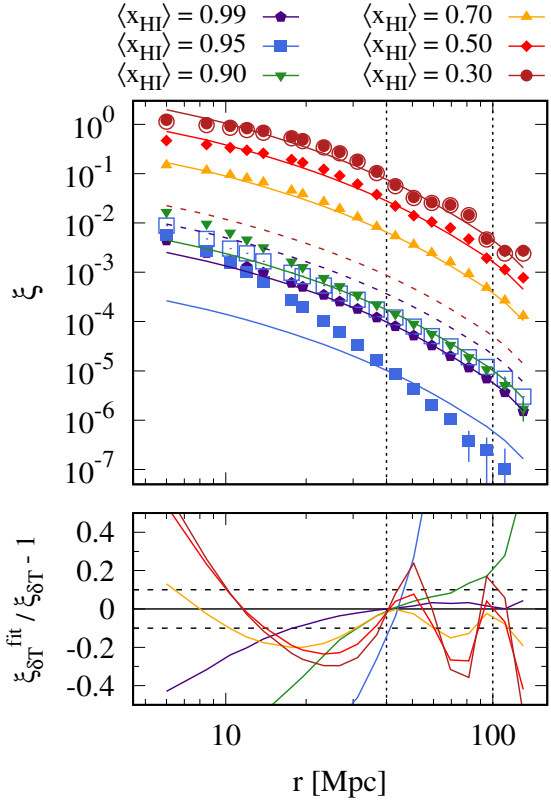}
\caption{
{\it Top}: The 2PCF of the 21cm brightness temperature $\delta T$
from model A without RSD (solid symbols) versus the scale $r$ at various
global neutral fractions \nf{} (see Fig. \ref{fig:nf-z} for the corresponding redshifts).
Open symbol show the 21cm 2PCF with RSD at \nf{}$=0.95$ and $0.3$. Dashed lines show
the matter 2PCF at \nf{}$=0.99$ and $0.3$. All measurements are shown as means over
$200$ random realizations of the simulation. Note that the $1 \sigma$ errors on the
21cm 2PCFs are smaller than the symbol size, while errors on the matter 2PCF are not
shown for clarity. Solid lines show fits to the 21cm 2PCF measurements without RSD,
based on the bias model prediction from equation (\ref{eq:2pc_bias}).
The fitting range ($40 - 100$ Mpc) is enclosed by vertical black short-dashed lines.
{\it Bottom}: Relative deviations between fits and measurements.
Deviations of $10\%$ are marked by horizontal black dashed lines.
\label{fig:2pc_b1}}
\end{figure}

To study the relation between the matter and 21cm 2PCF, we show both correlations together in Fig. \ref{fig:2pc_b1}.
The matter 2PCF, displayed at only two redshifts for clarity, shows the expected scale dependence,
while the amplitude increases weakly with decreasing redshifts (decreasing neutral fractions), due to the slow
growth of matter fluctuations at early times. We have verified that these measurements agree with theory predictions
based on the initial power spectrum of the simulation for $r \lesssim 100$ Mpc. Tests with simulations in
boxes with twice the side length (i.e. $1536$ Mpc) revealed that our mean 2PCF measurements at scales
above $r \gtrsim 100$ Mpc are significantly affected by the limited box size, since the latter causes
an artificial cut-off at large wavelength modes (see Appendix \ref{app:2PCF_boxsize}). We therefore
exclude 2PCF measurements for scales above $100$ Mpc from our analysis.

The 21cm 2PCF, displayed for model A without RSD as solid symbols at various redshifts in Fig. \ref{fig:2pc_b1},
shows a much stronger redshift evolution than the matter 2PCF. We find that the amplitude first decreases, between
\nf$= 0.99$ and $0.90$, before it increases strongly for lower redshifts with lower neutral fractions. At
large scales ($r \gtrsim 30$ Mpc) the shapes of the 21cm 2PCFs are very similar to those of the matter 2PCF,
while being shifted by a roughly constant factor. This latter behaviour can be expected from the bias model,
since equation (\ref{eq:quad_bias_model}) delivers at leading order in $\delta_m$
\begin{equation}
	\xi_{\delta T} \simeq b_1^2 \xi_m.
	\label{eq:2pc_bias}
\end{equation}
Interestingly, the 21cm 2PCF at \nf$= 0.95$ does not follow this trend, as its amplitude falls more steeply
with scale, indicating a very week clustering of the 21cm signal. This effect will be be discussed in Section \ref{sec:bias_2PCFvs3PCF}.
Note that alternatively to equation (\ref{eq:2pc_bias}), the relation between the $\xi_{\delta T}$ and $\xi_m$ can be derived 
by expanding the former in terms of some 2nd, 3rd, and 4th order statistics between neutral fraction and/or density fields 
and modelling them with physically motivated approximations.
However, those approximations can limit their accuracy \citep[e.g.][]{FZH04, Lidz07, Raste17}.
Furthermore, this approach becomes more challenging for statistics beyond the second order such as the 3PCF.

We fit the linear bias model for $\xi_{\delta T}$ from equation (\ref{eq:2pc_bias})
to our measurements between $40-100$ Mpc via a $\chi^2$ minimization,
as detailed in Appendix \ref{app:fitting}. Note that we can thereby only infer the absolute value
of the linear bias parameters, since it appears as $b_1^2$ in equation (\ref{eq:2pc_bias}).
However, we will see later on that $b_1$ can take positive as well as negative values.
The fits to equation (\ref{eq:2pc_bias}) are shown as solid lines in the top panel of Fig. \ref{fig:2pc_b1}.
They deviate from the measurements by $\lesssim 10\%$ ($\lesssim 30\%$) within the fitting range for
\nf{}$\gtrsim 0.7$ ($\gtrsim 0.3$), as shown in the bottom panel of Fig. \ref{fig:2pc_b1}. The strong
deviations for lower global neutral fractions, as well as for \nf{}$= 0.95$ indicate a breakdown
of the local bias model at these stages of reionization.
Deviations from the fits increase as well at small scales below $30$ Mpc, which correspond roughly
to the typical sizes of ionized regions, as shown in Fig. \ref{fig:bubble_sizes}. Deviations
at large scales above $100$ Mpc may partly
result from the aforementioned box size effect.

\begin{figure}
\centering\includegraphics[width=9.5 cm, angle=270]{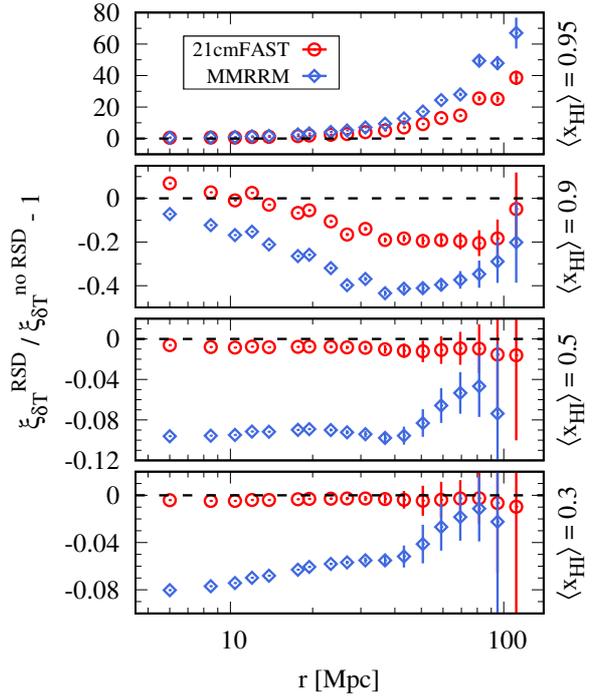}
\caption{Relative deviations between the 21cm 2PCF with and without RSD at
four redshifts with different global neutral fractions \nf. Red circles show
results based on a simplistic RSD model, implemented in \texttt{21cmFAST}. Blue diamonds
show results for a more realistic RSD model from \citet{Mao12}.
\label{fig:2pc_RSD}}
\end{figure}

Note that the 21cm 2PCFs which we use for the fits do not include
RSDs, which will not be the case in future observational data sets. In order to obtain a rough
insight into how RSDs would affect the 21cm 2PCF, we show results from simulations
with RSDs from the MMRRM model as open symbols in Fig. \ref{fig:2pc_b1}. The results
show that the RSDs cause a strong increase of the 21cm 2PCF at large scales and early time (\nf$=0.95$),
while the latter takes a similar shape as the matter 2PCF. At later times (\nf$=0.3$) the effect of RSDs
decreases strongly.
A more detailed picture can be obtained from Fig. \ref{fig:2pc_RSD}, which shows the relative
deviations between the 21cm 2PCF for model A with and without RSDs. We compare results
from the MMRRM model to those from a simplified RSD model, implemented in \texttt{21cmFAST}.

The MMRRM model predicts a strong increase of the 21cm 2PCF by up to a factor of $60$ at large scales for
\nf$=0.95$ and a strong decrease of up to $50 \%$ at \nf$=0.9$ due to RSD. At later times (\nf{}$\le 0.5$),
the decrease is strongest at small scales ($r\lesssim 10$ Mpc) with $\lesssim 10\%$ shift of
the amplitude and becomes insignificant at $r \gtrsim 60$ Mpc.
The simplified RSD model from \texttt{21cmFAST} tends to underpredict the RSD effect on the 2PCF,
in particular at late times. However, at early times, where RSD effects are most significant,
the approximate implementation works reasonably well.  We conclude that RSDs need to be taken
properly into account, when interpreting future observational data sets, but we focus in this work
on simulations without RSD to develop a physical understanding of how the matter and 21cm correlations
are related to each other.

\subsection{3-point correlation functions (3PCF)}\label{sec:3pc}
The 3PCF is defined analogously to the 2PCF as
\begin{equation}
		\zeta(r_{1}, r_{2}, r_{3}) \equiv
		\langle \delta_1 \delta_2 \delta_3 \rangle(r_{1}, r_{2}, r_{3}).
		\label{eq:3pc_def}
\end{equation}
The positions of the fluctuations $\delta_{i}=\delta({\bf x}_{i})$ form a triangle, with three legs of the sizes $r_{1} = |{\bf x}_2-{\bf x}_1|$, 
$r_{2} = |{\bf x}_3-{\bf x}_2|$, and $r_{3} = |{\bf x}_1-{\bf x}_3|$.
Similar to the 2PCF, $\langle ... \rangle$ denotes the average over all possible triangle orientations and translations.
We denote the 3PCFs of the matter and 21cm brightness temperature fluctuations (referred to as ``21cm 3PCF'') as $\zeta_m$
and $\zeta_{\delta T}$, respectively. Our 3PCF measurements
are performed on the same type of grid on which we compute the 2PCF. To search for triplets of
grid cells, we employ the algorithm described by \cite{BaGaz02}. This algorithm delivers measurements for triangles
with two fixed legs ($r_1, r_2$) and varying leg sizes $r_3$. The cell size
is set to $6$ or $12$ Mpc for small and large triangles, respectively, to optimize the computation time.
We have verified for the triangle configurations $(r_1, r_2) = (72, 36)$ and $(48, 48)$ Mpc, that the effect
of the cell size (or smoothing scale) on the 3PCF measurements is within the $1\sigma$ errors, except for triangles for which $r_3$
is comparable to the cell size \citep[see also][]{Hoffmann17}. In general, the 3PCF depends on the smoothing scale,
even at large scales, while we expect this smoothing effect on our bias measurements described below to be small,
based the aforementioned test.
%
\begin{figure*}
\centering\includegraphics[width=10 cm, angle=270]{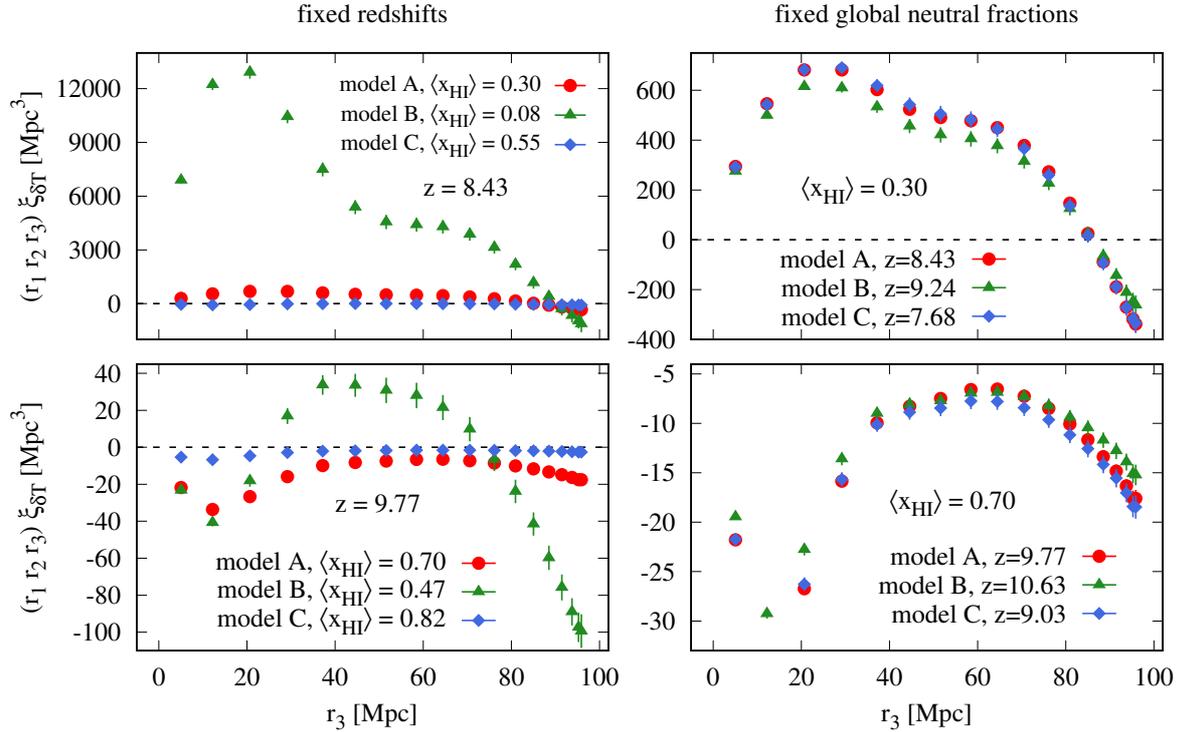}
\caption{21cm 3PCF for triangles with two fixed legs of $(r_1, r_2) = (48, 48)$ Mpc
versus the size of the third leg $r_3$. The smallest and largest $r_3$ values correspond
to collapsed and relaxed triangles with opening angles of $0$ and $180$ degrees, respectively. 
Results are shown for the three EoR models defined in Table \ref{table:EOR_param} without RSD at
fixed redshifts $z$ and fixed volume weighted global neutral fractions \nf\ (left and right panels respectively).
The 3PCFs are multiplied by the product $(r_1r_2r_3)$ to facilitate the visual model comparison at different scales,
which modifies the overall shape of the measurements. Note that $1\sigma$ errors can be smaller than
the symbols.}
\label{fig:3pc_EORmodels}
\end{figure*}

We show measurements of the 21cm 3PCF for the three EoR models, summarized in Table \ref{table:EOR_param},
without RSDs at the redshifts $z=8.43$ and $9.77$ in the left panels of Fig. \ref{fig:3pc_EORmodels}. The measurements
are done using triangles with $(r_1, r_2) = (48, 48)$ Mpc and are displayed versus the 
third triangle leg $r_3$, for $18$ corresponding opening angles $\alpha$ (the angle between $r_1$ and $r_2$)
between $0$ and $180$ degrees. These measurements reveal a high sensitivity of the 21cm 3PCF to variations of the
EoR model parameters. The strong change of the 3PCF amplitude as well as its shape stand in contrast to our 2PCF
measurements, for which we found changes due to variations of EoR parameters to be well described by a simple shift
of the amplitude (Fig. \ref{fig:2pc_EORmodels}). This finding illustrates that 3PCF contains 
additional information with respect to the 2PCF, which may be used to tighten constraints on EoR models with future observations. In the right panel of Fig. \ref{fig:3pc_EORmodels}
we compare the 21cm 3PCF from different models at fixed global neutral fractions. We find a relatively
weak ($\lesssim 10 \%$) dependence of the measurements on the model parameters in that case, compared to the change
with redshift. This result lines up with our finding for the 2PCF from the Fig. \ref{fig:2pc_EORmodels} and may be
explained with the same argument outlined in Section \ref{sec:2pc}.

\begin{figure*}
\centering\includegraphics[width=9 cm, angle=270]{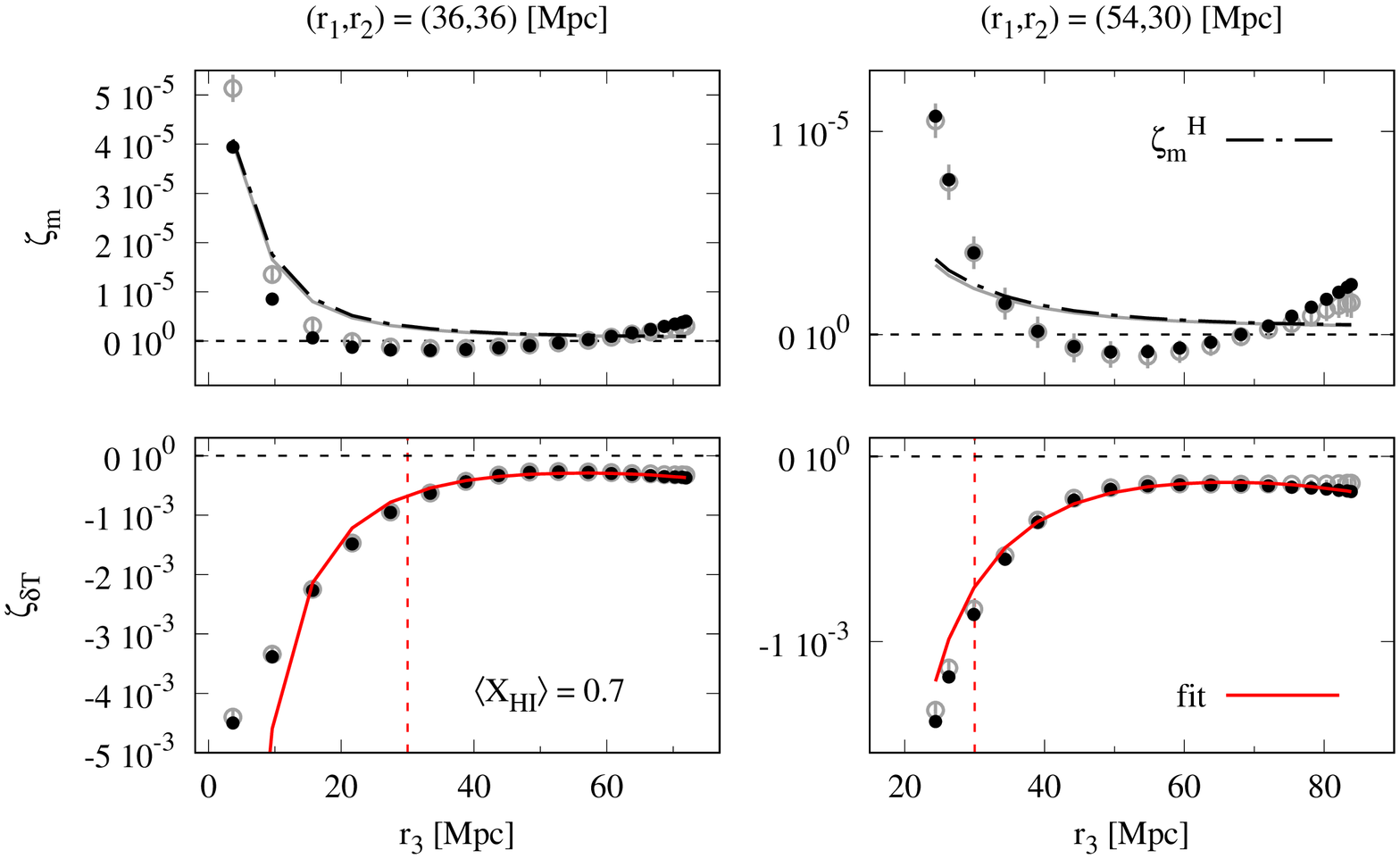}
\caption{The 21cm 3PCF for two triangle configurations, defined by the fixed legs $(r_{1}, r_{2})$ versus the size 
    of the third leg $r_3$, at redshift $z=9.77$ (\nf$=0.7$). Top panels: measurements of the matter 3PCF (black dots),
    and the hierarchical matter 3PCF $\zeta^H \equiv \xi_{12}\xi_{13} + {\rm 2\,\,permutations}$ (black dashed-dotted lines). 
    Bottom panels: the corresponding 21cm 3PCF (black dots), with fits to leading order prediction from the bias model
    in equation (\ref{eq:3pc_bias}) (red solid line). The minimum scale used for the fit, i.e. $30$ Mpc, is marked
    by a vertical red dashed line. Note that the $1 \sigma$ errors are smaller than the symbol size. Results
    based on simulations with \texttt{L-PICOLA} matter distributions are shown as grey open dots and solid lines for the 3PCF and
    hierarchical matter 3PCF respectively.
}
\label{fig:3pc_configs}
\end{figure*}

We continue our study by investigating the relation between the 3PCF of the 21cm and the matter field
in Fig. \ref{fig:3pc_configs} for two triangle configurations with fixed legs $(r_1, r_2) = (36, 36)$ and $(54, 30)$ Mpc
versus $r_3$ at redshift $z=9.77$ (\nf$=0.7$).
The matter 3PCF, displayed in the top panels, has positive values for collapsed and relaxed triangles (small and large $r_3$ respectively),
while being negative for half open triangles. This typical shape is a signature of the filamentary structure of the cosmic web
\citep[e.g.][]{Bernardeau02}, and hence provides additional information about the dark matter distribution to which one- or
two-point statistics are not sensitive.

These results are compared to the matter 3PCF from the \texttt{L-PICOLA} runs, which are shown as grey
open circles in the same figure. The shape of the matter 3PCF from both types of simulations are slightly different
from each other. However, these differences are small compared to the overall change of the 3PCF with triangle
configuration and lie within the $1\sigma$ uncertainties for triangles with $r_3 > 20$ Mpc.
The corresponding measurements for the 21cm 3PCF from model A without RSD are shown in the bottom panels of Fig. \ref{fig:3pc_configs}.
We compare also these measurements to results based on the \texttt{L-PICOLA} runs and find again only small deviations
compared to the dependence of the amplitude on triangle configuration and EoR model.

Interestingly these 21cm 3PCF measurements are negative for all triangles, while their overall shape appears to be similar to the matter 3PCF.
This indicates that there is a physical relation between both statistics. In analogy to the 2PCF, this relation can be
approximated by a perturbative expansion of the 21cm 3PCF, using the
bias model. From equation (\ref{eq:quad_bias_model}), one finds, at leading order
\begin{equation}
	\zeta_{\delta  T} \simeq b_1^3\zeta_m + b_1^2b_2\zeta_m^H
	\label{eq:3pc_bias}
\end{equation}
\citep{FryGa93}. The hierarchical three-point correlation, $\zeta^H \equiv \xi_{12}\xi_{13} + \xi_{21}\xi_{23} + \xi_{31}\xi_{32}$, 
with $\xi_{ij} = \xi(|{\bf x}_i-{\bf x}_j|)$, is
shown as dashed-dotted lines in the top panels of Fig. \ref{fig:3pc_configs}.

We derive $b_1$ and $b_2$ from the 3PCF by fitting equation (\ref{eq:3pc_bias}) to our measurements. The fits,
shown as red line in the bottom panels of Fig. \ref{fig:3pc_configs}, are obtained from
a $\chi^2$ minimization. The $\chi^2$ values are computed via a Singular Value Decomposition (SVD)
of the covariance matrix (shown in Fig. \ref{fig:3PCF_cov}), as suggested by \citet{GazSco05} to reduce
the impact of noise in the covariance on our fits (see Appendix \ref{app:fitting} for details).
We fit the model only to measurements from triangles, for which all legs are in the range $30-100$ Mpc.
This selection is motivated by our 2PCF results, which showed that the bias model prediction breaks down at smaller scales,
while measurements at large scales may be affected by the limited box size (see Section \ref{sec:2pc}).

Our fits are in reasonable agreement with the measurements within the fitting range, which shows
that the quadratic bias model can describe the dominating part of the 21cm 3PCF signal at large scales.
However, the fits deviate strongly from our measurements at small scales, below the minimum
fitting scale. This indicates a break down of the bias model, according to which a given set of fitted bias
parameters should describe the 21cm 3PCF at all scales.

\begin{figure}
\centering\includegraphics[width=15 cm, angle=270]{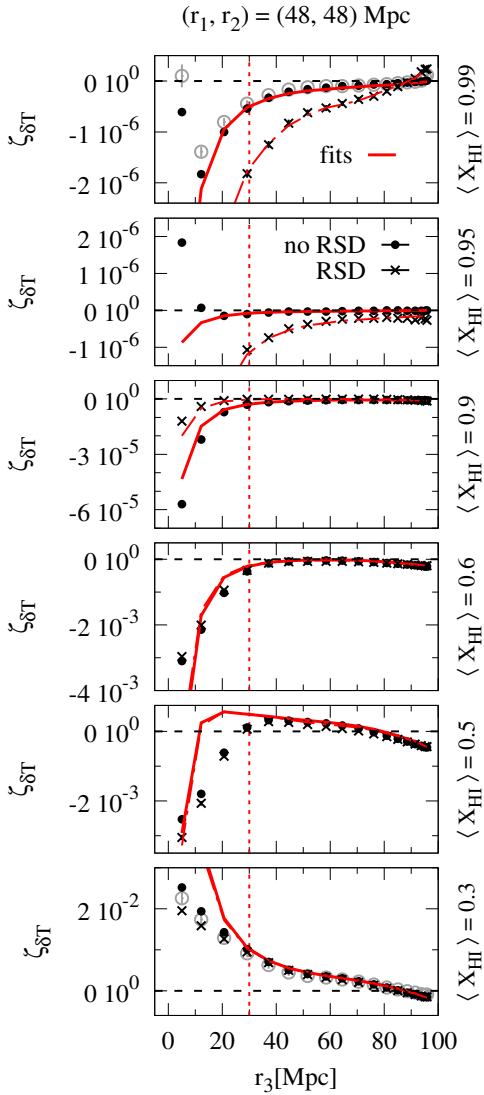}
\caption{21cm 3PCF for triangles with $r_1=r_2=48$ Mpc for $18$ opening angles between
    $0$ and $180$ degrees versus the corresponding leg size $r_3$.
    Each panel shows measurements for redshifts with different global neutral fractions \nf{}. Black dots and crosses
    show mean measurements from $200$ realizations for simulations without RSD and with RSD from
    the MMRRM model, respectively. Note that the $1\sigma$ errors are smaller than the
    symbol size. Results based on \texttt{L-PICOLA} matter distributions are shown as grey open dots
    for \nf{}$=0.99$ and $0.3$.
    Solid and dashed red lines show fits to the leading order prediction from the
    bias model from equation (\ref{eq:3pc_bias}) to results without and with RSDs respectively.
    The fits are performed for triangles with $r_3>30$ Mpc, as marked by vertical red dotted lines.}
\label{fig:3pc_z}
\end{figure}

The redshift evolution of the 21cm 3PCF is shown for triangles with $(r_1, r_2) = (48, 48)$ Mpc
versus $r_3$ in Fig. \ref{fig:3pc_z}. We use again model A, while we now show measurements
for simulations without RSD as well as with RSD from the MMRRM model.
Focusing first on results without RSD, we find that on large scales and at early times
($r_3\gtrsim30$ Mpc, \nf{}$\gtrsim 0.6$) the 21cm 3PCF has a negative amplitude and a positive slope for this
triangle configuration. At later times the large-scale amplitude becomes positive, while the slope becomes negative at large scales.
Again, this behaviour is well described by the fits to equation (\ref{eq:3pc_bias}) on
scales above $30$ Mpc, while we find strong deviations between fits and measurements
at smaller scales. An interesting feature at small scales is the local minimum of the
3PCF without RSD at $r_3 \simeq 10$ and $20$ Mpc for \nf$= 0.99$ and $0.95$, respectively.
This feature may be associated with the a small scale increase
of the 21cm bispectrum, predicted by \citet{Majumdar17} for randomly distributed ionized bubbles.
These authors also found a similar feature in the 21cm bispectrum measurements from simulations.
However, more detailed studies are needed for understanding this effect.

As in Fig. \ref{fig:3pc_configs} we compare in Fig. \ref{fig:3pc_z} our main results for the 21cm 3PCF based on the Zel'dovich
approximation to those from the \texttt{L-PICOLA} runs, which are shown as grey open circles at the
highest and lowest redshifts with \nf{}$=0.99$ and $0.3$ respectively. We find again that the
differences between results from both types of simulations are small, in particular at large scales,
compared to the change of the amplitude with triangle configurations and redshift. We therefore do not
expect the conclusions of this investigation to be significantly affected by the usage of the Zel'dovich aproximation.

The 21cm 3PCFs from simulations with RSD appear to be very similar to those from simulations
without RSD at large scales and late times ($r_3\gtrsim30$ Mpc, \nf{}$\lesssim 0.9$), while at
smaller scales as well as at earlier times results from both cases deviate more significantly. The RSD results
can also be well fitted with the quadratic bias at large scales. However, the RSD change the physical
meaning of the fitted bias parameters, since in that case the 21cm 3PCF does not only depend on the
relation between the matter field and the hydrogen distributions, but also on the hydrogen velocity.

\begin{figure}
\centering\includegraphics[width=10.2 cm, angle=270]{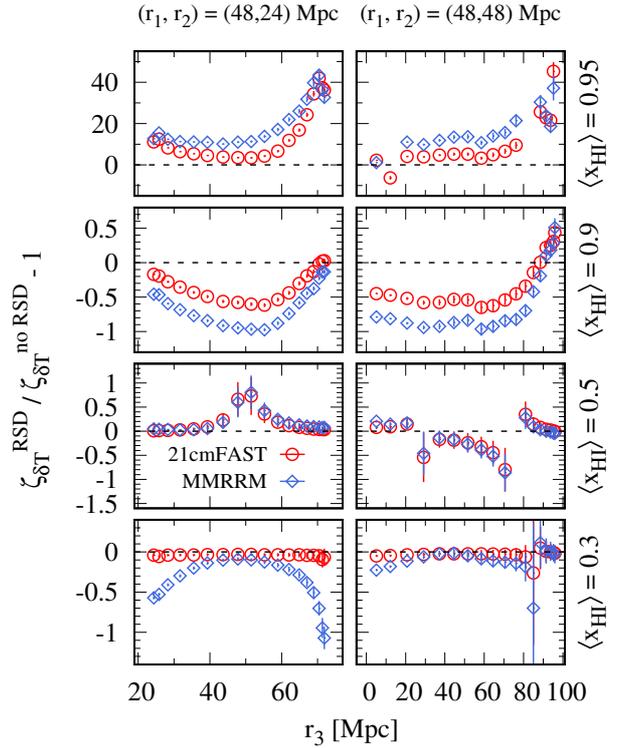}
\caption{Analogous to Fig. \ref{fig:2pc_RSD} for the 21cm 3PCF.
Left and right panels show results for two different triangle configurations,
characterized by the fixed legs $r_1$ and $r_2$ versus the size of the third leg $r_3$.}
\label{fig:3pc_RSD}
\end{figure}

In Fig. \ref{fig:3pc_RSD} we compare the relative deviations between the 21cm 3PCF
from simulations with model A with and without RSD for two triangle configurations with
fixed legs of $(r_1, r_2) = (48, 24)$ and $(48,48)$ Mpc at different $r_3$ sizes and
different global neutral fractions. We find the RSD effect to depend strongly on the
specific triangle scale and configurations, as well as on redshift. For the MMRRM model
the changes in the amplitude vary between $\simeq 10\%$ (e.g. for $r_1=r_2=r_3=48$ Mpc at
\nf$=0.3$) and a factor of $40$ (for the largest $r_3$ values at \nf$=0.95$). The relative
changes of the 21cm 3PCF due to RSD are overall stronger, than those which we found for the
21cm 2PCF in Fig. \ref{fig:2pc_RSD}. However, note that even strong changes in the
3PCF amplitude can lead to relatively small changes in the bias parameters (see
Section \ref{sec:bias_comparison}), since $b_1$ enters equation (\ref{eq:3pc_bias})
with up to the power of three. The absolute change of the 3PCF amplitude from simulations
with the simplified RSD model, implemented in \texttt{21cmFAST} is lower than predicted
by the more physical MMRRM model, which lines up with our 2PCF results from Fig. \ref{fig:2pc_RSD}.

\begin{figure*}
\centering\includegraphics[width=4.9 cm, angle=270]{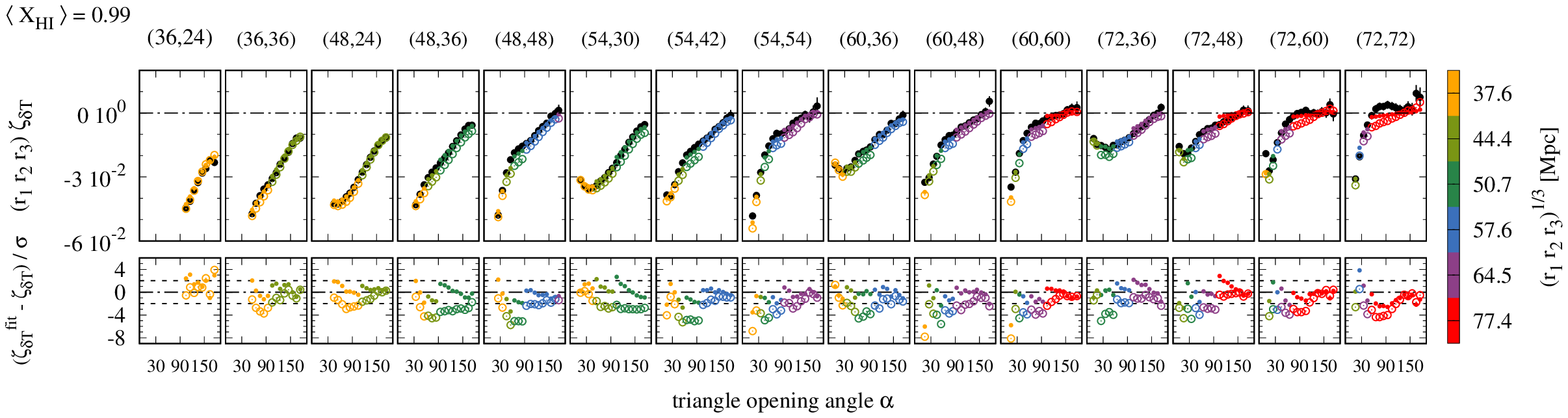}
\centering\includegraphics[width=4.9 cm, angle=270]{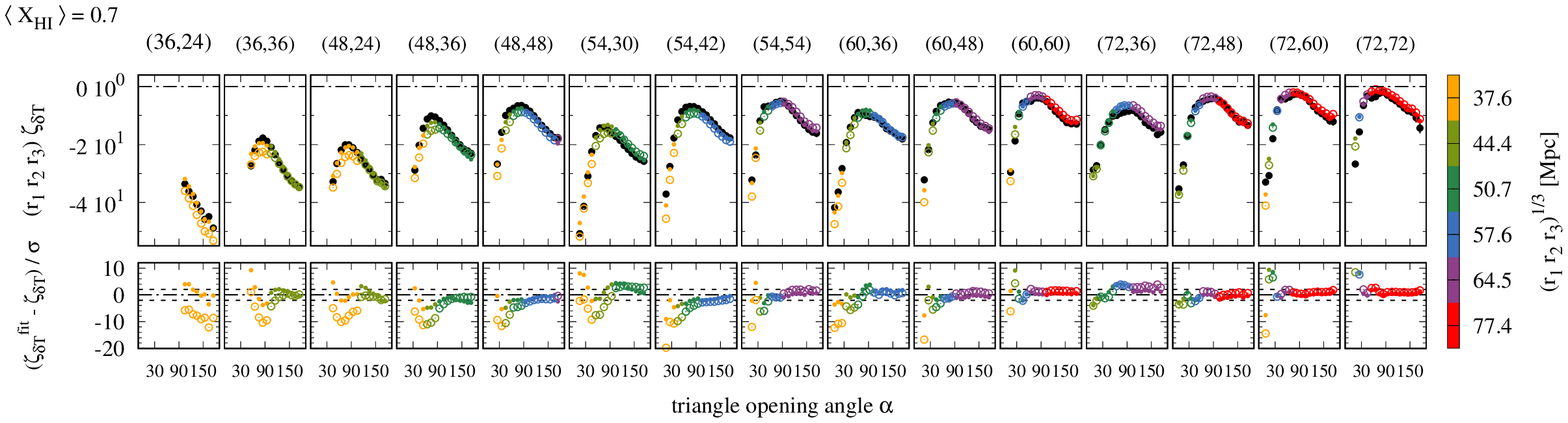}
\caption{
    Measurements of the 21cm 3PCF from model A without RSD for the global neutral fractions \nf{}$=0.99$ and $0.7$ (black dots in the
    top and bottom panels respectively). Each column shows results from one triangle configuration
    versus the triangle opening angle. The configurations are defined by the fixed legs $(r_1, r_2)$,
    as indicated in Mpc on the top. The 3PCFs are multiplied by the scale $(r_1 r_2 r_3)$ to
    conveniently display all measurements on a common y-axis (the latter has therefore the unit $\text{Mpc}^3$).
    Coloured open circles show fits to the leading order bias models prediction. Coloured dots show fits, without
    taking off-diagonal elements of the error-covariance into account. The fits are performed
    separately for triangles in $6$ different $(r_1r_2r_3)^{1/3}$ scale bins. Each bin
    contains $40$ triangles, which are marked by the same colour. The mean scale of each bin is
    displayed in Mpc next to the colour bar on the right. The lower sub-panels show the significance of the
    deviation between fits and measurements in units of the $1\sigma$ measurement errors. Horizontal
    short-dashed lines mark $2\sigma$ deviations.
}
\label{fig:3pc_allconfigs}
\end{figure*}

We now aim at testing the quadratic bias model prediction for the  21cm 3PCF from model A without RSD
on a wider range of scales by jointly fitting larger sets of triangles. In Fig. \ref{fig:3pc_allconfigs}
we show 3PCF measurements from $15$ triangle configurations, defined by the fixed legs $(r_1, r_2)$ at the
redshifts $z=13.49$ and $9.77$, with \nf{}$=0.99$ and $0.7$ respectively. The 3PCF
is multiplied by $(r_1r_2r_3)$ to decrease the
strong scale dependence of the signal, simplifying a visual comparison
of the different measurements. Note that this multiplication
modifies the overall shape of the 3PCF, which can be seen by comparing
results for $r_1=r_2=48$ Mpc to the results from Fig. \ref{fig:3pc_z}.
In order to verify the scale dependence of the 3PCF bias model fitting performance and the resulting bias parameters,
we fit the measurements in $6$ triangle scale bins, as marked by colour in Fig. \ref{fig:3pc_allconfigs}. The
triangle scale is thereby defined as $(r_1r_2r_3)^{1/3}$, while each scale bin contains $40$ triangles.
We find deviations between fits and measurements of roughly $2 \sigma$
for the largest triangle scale bin with $\langle (r_1r_2r_3)^{1/3} \rangle = 77.4$ Mpc
at both redshifts. For the smallest scale bin with $\langle (r_1r_2r_3)^{1/3} \rangle = 37.6$ Mpc,
deviations tend to be more significant, with up to $4$ ($10$) $\sigma$ for \nf$=0.99$ ($0.7$). These values can correspond
to relative deviations of more than $50\%$ for \nf{}$\lesssim0.6$, while we find $\simeq 20 \%$ deviations for
$(r_1 r_2 r_3)^{1/3} \lesssim 60$ Mpc and higher global neutral fractions (see Fig. \ref{fig:3PCF_reldiff}).
The latter figure also suggests that the lower significance of deviations at large scales
might be attributed to an increase of the measurement errors with scale. However, one could
also expect that the perturbative approximations incorporated in the bias model should work better at larger scales,
where non-linearities are smaller. To verify the impact of the covariance estimate on the fits, we neglect
off-diagonal elements in the covariance for computing the $\chi^2$ deviations in eq. (\ref{eq:chisq}).
The resulting fits, shown as colored dots in Fig. \ref{fig:3pc_allconfigs}, indicate that
that these off-diagonal covariance matrix elements can have a strong impact on the fits and should not be ignored.
It is therefore important to reduce noise in the covariance estimates, which we attempt by using the aforementioned
SVD technique and a relatively large number of realizations.

\begin{figure}
\centering\includegraphics[width=12 cm, angle=270]{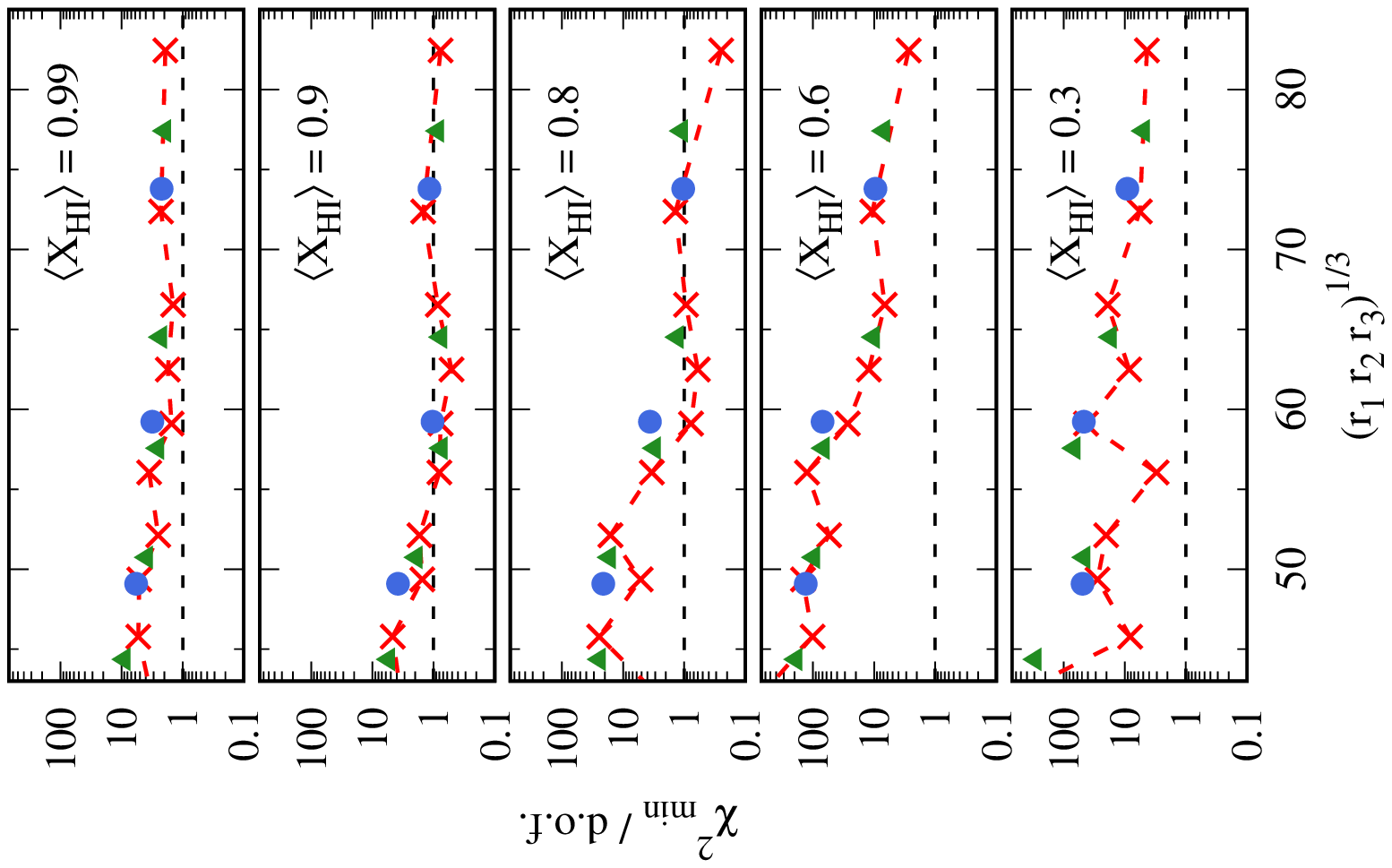}
\caption{
    Minimum $\chi^2$ per degree of freedom, quantifying the fitting performance of the quadratic bias model
    prediction for the 21cm 3PCF from equation (\ref{eq:3pc_bias}) for different global neutral fraction \nf{}.
    The model was fitted to 3PCF measurements for model A without RSD from triangles in bins,
    defined by the triangle scale $(r_1r_2r_2)^{1/3}$. The mean scale per bin is shown on the x-axis.
    Results are shown for $20$, $40$ and $60$ triangles per bin
    as red, green and blue symbols respectively, while in the first case the symbols are connected with lines
    to highlight the scale dependence. The degree of freedom is the number of modes in the 3PCF covariance
    for a particular set of triangles, which is above the noise limit and used for the $\chi^2$ computation.
}
\label{fig:chisq-r_Zfit}
\end{figure}

The goodness of these fits is shown as the minimum $\chi^2$ per degree
of freedom ($d.o.f$) in Fig. \ref{fig:chisq-r_Zfit} versus the mean triangle scale per bin for model
A without RSD at various volume weighted neutral global fractions. The degree of freedom is here the
number of modes in the covariance matrix with singular values above the shot-noise limit, from which
we compute the $\chi^2$ values (see Fig. \ref{fig:3PCF_covSV}).
In addition to the $\chi^2/d.o.f$ values based on fits in scale bins with $40$ triangles,
we also show results for smaller and larger bins with $20$ and $60$ triangles, respectively, verifying the robustness of our results.
We find that for scales above $60$ Mpc, the $\chi^2/d.o.f$ values range between $1$ and $10$, independently from the binning.
Given our small measurement errors, which result from the large total simulation volume
of the $200$ realizations, the goodness of fit values indicate a reasonable agreement between model
and measurements for \nf{}$\ge0.6$. The $\chi^2/d.o.f$ values strongly increase for lower global 
eutral fractions, which is another indication for the break down of the leading order 3PCF bias model.
The increase of the $\chi^2/d.o.f$ values at small scales might be attributed to the same effect.
However, the smaller measurement errors at small scales also increase the significance of deviations
between measurements and model fits, as we already concluded from Fig. \ref{fig:3pc_allconfigs}.

\section{Bias comparison}\label{sec:bias_comparison}

In the previous section, we found that the leading order bias expansions of
the 21cm 2PCF and 3PCF provide reasonable fits to the measurements in our simulations,
in particular at early times of reionization and at large scales.
In this section, we will focus on this regime, studying the bias parameters obtained from these fits
in two ways. We start by verifying the consistency of the bias parameters,
obtained from the 2PCF and the 3PCF fits in Section \ref{sec:bias_2PCFvs3PCF}.
In Section \ref{sec:scatter_plot} we will test if the deterministic quadratic bias model
is in agreement with the relation between the
fluctuations of \dT{} and matter, which we can measure directly in the simulations.

\subsection{Bias from correlation functions}\label{sec:bias_2PCFvs3PCF}

\begin{figure}
\centering\includegraphics[width=7 cm, angle=270]{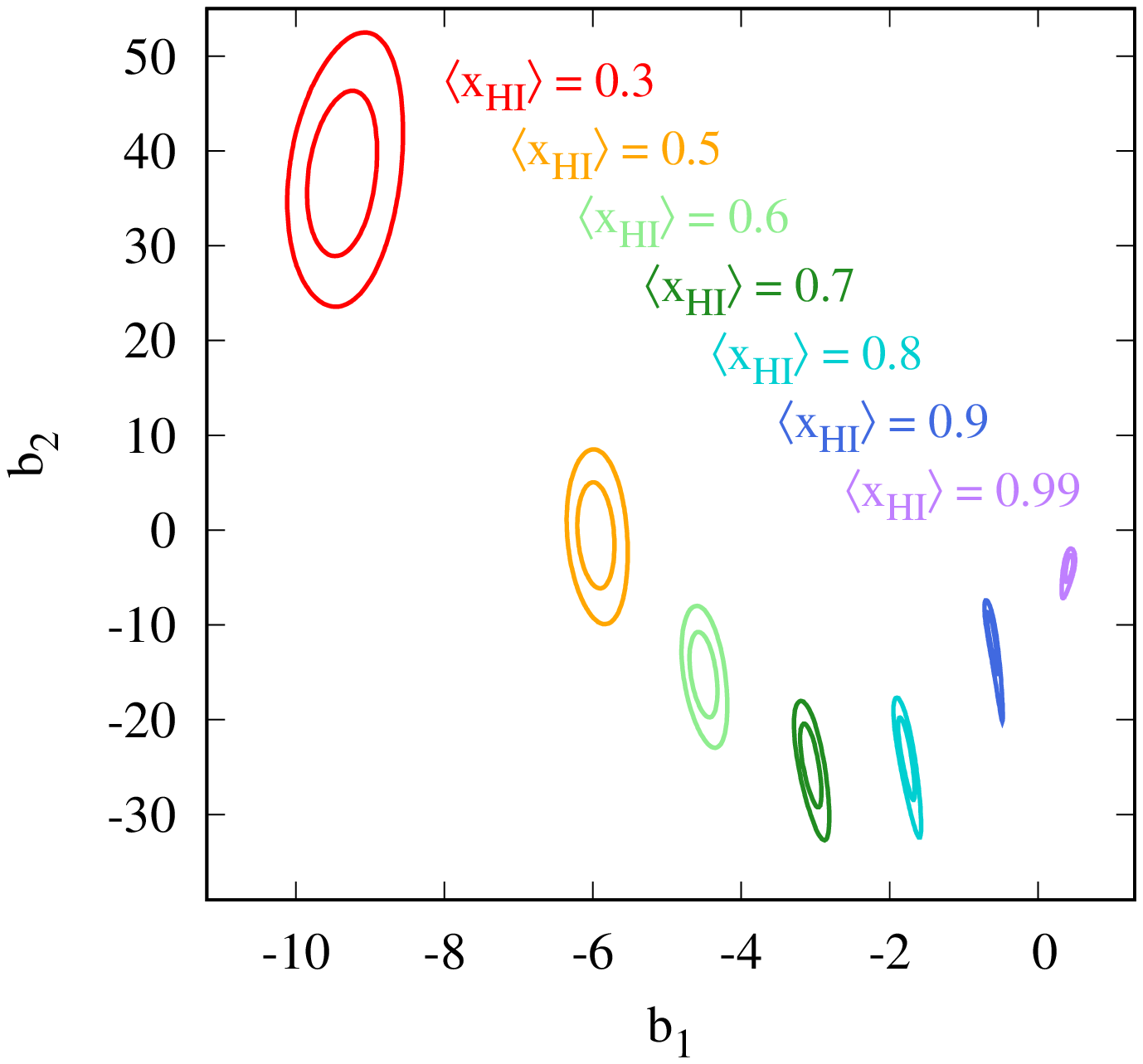}
\caption{
    Confidence levels of the linear and quadratic bias parameters from fits of
    the quadratic bias model prediction for the 21cm 3PCF to measurements
    in model A simulations without RSD for different neutral fractions \nf{}. Contours enclose regions with
    $\chi^2-\chi^2_{min} \lesssim 2.30$ and $\lesssim  6.17$, which corresponds to $68.3$ and
    $95.4 \%$ probabilities for a Gaussian distribution. The fits are based on $40$ triangles
    from our sample at the largest $(r_1r_2r_3)^{1/3}$ bin, i.e.\ $68.05 \le (r_1r_2r_3)^{1/3} \le 90.68$ Mpc 
    (see Fig. \ref{fig:3pc_allconfigs}). }
\label{fig:b1b2_chisq}
\end{figure}

Examples of our 3PCF bias constraints are shown in Fig. \ref{fig:b1b2_chisq}
for our fiducial model A without RSD at different volume weighted global neutral
fractions as $68.3\%$ and $95.5\%$ confidence regions in the $b_1-b_2$ parameter space. These results are
obtained from $40$ triangles with an average scale of $\langle (r_1r_2r_3)^{1/3} \rangle \simeq 77.4$ Mpc,
shown as the largest scale bin in Fig. \ref{fig:3pc_allconfigs}.
We find the linear bias $b_1$ to be positive for \nf$=0.99$ and negative
at later times, with values between $-5 \lesssim b_1 \lesssim 1$.
The initially positive values of our $b_1$ measurements originate from the fact that
the HI distribution follows the matter density before reionization
starts. Later on, the first ionized regions with zero 21cm emission form around the
largest matter overdensity regions (see Fig. \ref{fig:sim}), leading to an anti-correlation
between the 21cm signal and the matter fluctuations, as described by a negative
bias (see Section \ref{sec:scatter_plot} for a more detailed discussion).
The change of $b_1$ from positive to negative values with time
is consistent with the decrease of the 2PCF amplitude at early times and its increase
at later times (Fig. \ref{fig:2pc_b1}), since the latter depends on $b_1^2$.
The transition around $b_1\simeq0$ manifests itself in the low 2PCF amplitude at \nf{}$=0.95$,
shown in the same figure.
The quadratic bias $b_2$ is negative for \nf{}$\gtrsim0.5$ and positive at later times with lower global neutral fractions.
It reaches larger absolute amplitudes than $b_1$, in the rage $-25 \lesssim b_2 \lesssim 40$
with a minimum of $b_2$ at \nf$\simeq0.7$. The degeneracy between the two parameters is relatively weak,
while the uncertainty on $b_2$ is significantly larger than for $b_1$.
The latter result might be attributed to the fact that $\zeta_{\delta T}$ is more sensitive
to changes in $b_1$ than in $b_2$ since these parameters appear in equation (\ref{eq:3pc_bias}) at
up to third and linear order respectively.
%

\begin{figure*}
\centering\includegraphics[width=9 cm, angle=270]{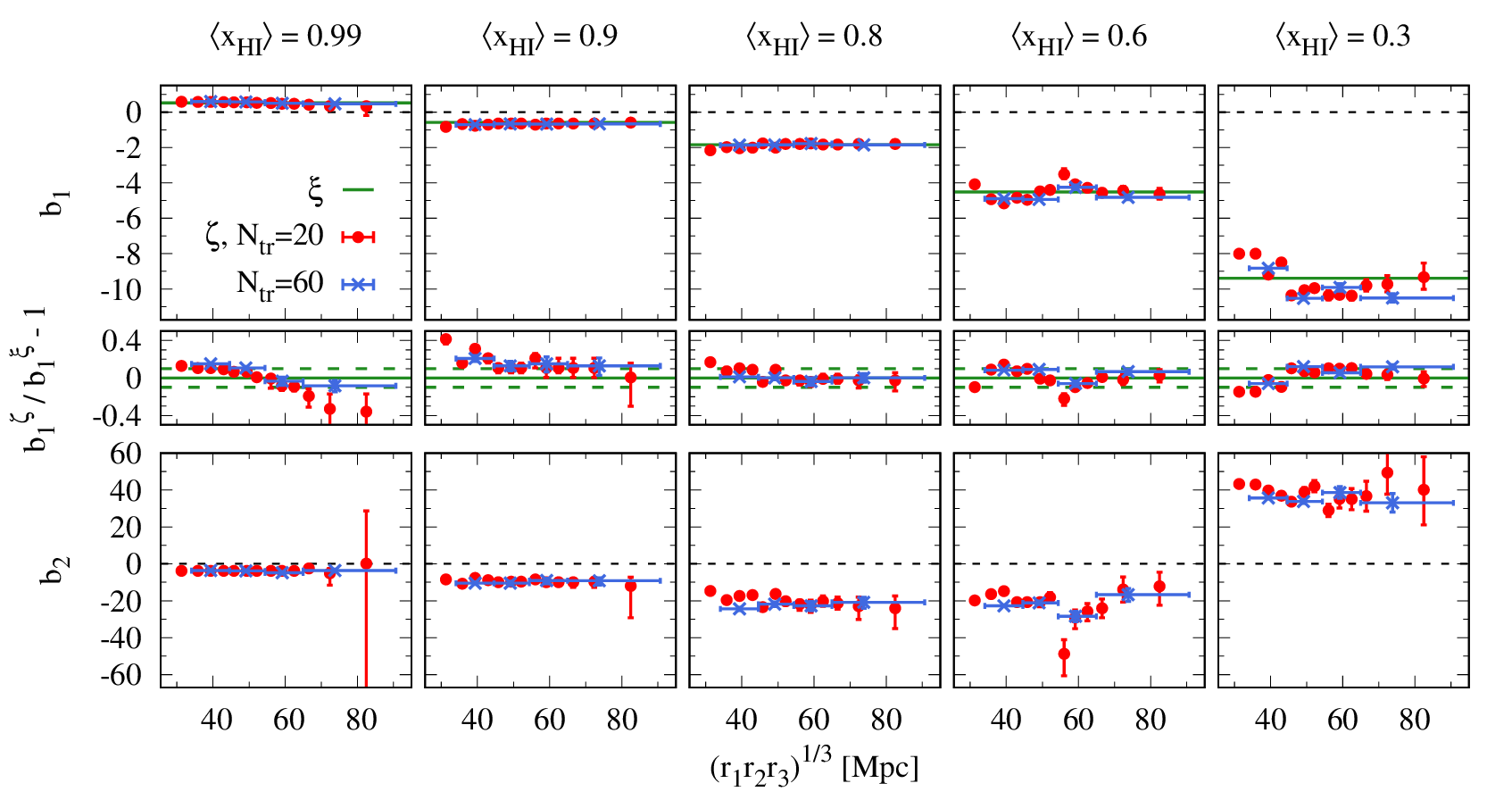}
\caption{
    {\it Top}: Linear bias parameters, measured from fits to the 21cm 3PCF from model A without RSD
    using triangles in bins of the triangles scale $(r_1r_2r_3)^{1/3}$, as
    shown in Fig. \ref{fig:3pc_allconfigs}. Results are shown for scale bins
    with $20$ and $60$ triangles as red and blue dots respectively. The linear
    bias from fits to the 2PCF from Fig. \ref{fig:2pc_b1} is shown as solid green
    line with $10\%$ deviations marked by green dashed lines. Each column
    shows results for different global neutral fractions \nf.
    {\it Center}: Relative difference between $b_1$ from 2PCF and 3PCF fits.
    {\it Bottom}: Corresponding measurements of the quadratic bias.
}
\label{fig:b1b2_r}
\end{figure*}

To verify the scale dependence of our 3PCF bias measurements, we display $b_1$ and $b_2$ in
Fig. \ref{fig:b1b2_r} for model A without RSD, derived from triangles in different $(r_1r_2r_3)^{1/3}$
scale bins versus the mean scale in each scale bin. Results are shown for narrow and broad scale bins,
containing $20$ and $60$ triangles, respectively, at redshifts with global neutral fractions between
\nf$=0.99$ and $0.3$. Note that we take all triangles into account. However, excluding triangles
for which one of the legs is smaller than $30$ Mpc (as done for the fits, shown in
Fig. \ref{fig:3pc_configs} and \ref{fig:3pc_z}) affects only results from the smallest scale bins.
We find only a weak scale dependence of the bias parameters for redshifts with \nf$> 0.6$.
At lower redshifts with lower neutral fractions, we find an increasing scale dependence with $\simeq 20\%$ ($\gtrsim 50\%$)
variations of $b_1$ ($b_2$), with respect to the average over all scales.
These variations indicate an increasing inaccuracy of the bias model as reionization
progresses, which is consistent with our 2PCF results from Fig. \ref{fig:2pc_b1}. Overall, results
from small and large-scale bins are consistent, indicating that they are not strongly
affected by noise in the covariance estimation.

The different $b_1$ measurements from the 3PCF fits, shown in Fig. \ref{fig:b1b2_r},
are compared in the same figure to the $b_1$ measurements from the 2PCF fits between
$40$ and $100$ Mpc (see Section \ref{sec:2pc}). Note that the sign for the latter is set by hand,
since the 2PCF only delivers constraints on the absolute value of $b_1$.
We find that, in general, the linear bias from the 3PCF is in $10\%$ agreement with 2PCF fits
which corresponds roughly to the inaccuracy of the 2PCF fits from Fig. \ref{fig:2pc_b1}.

\begin{figure}
\centering\includegraphics[width=11 cm, angle=270]{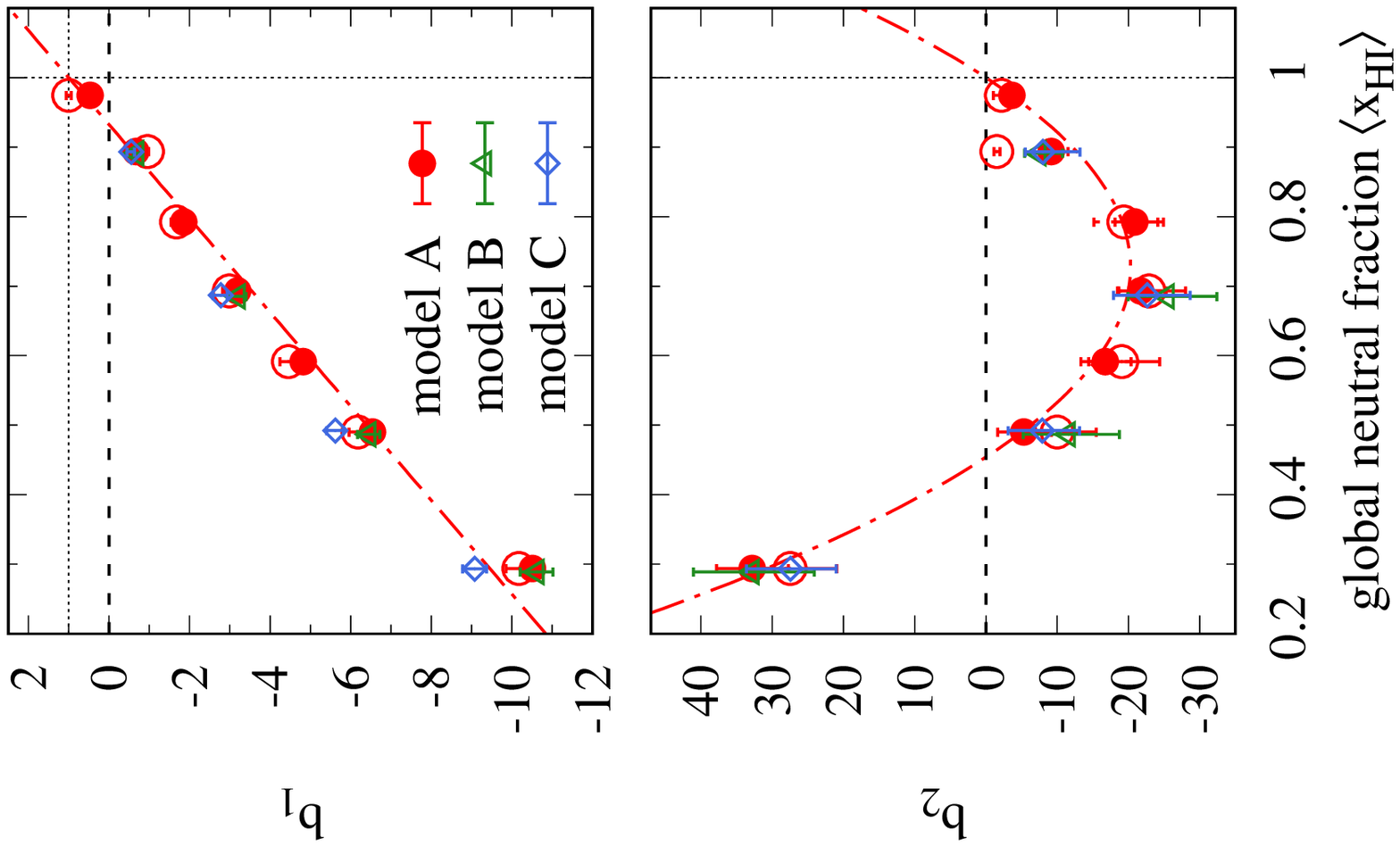}
\caption{
    Linear and quadratic bias parameters (top and bottom panels, respectively),
    derived from the 21cm 3PCF versus the global volume weighted neutral fraction $\langle x_{\rm HI} \rangle$.
    Red filled circles, green triangles and blue diamonds show measurements derived from model A, B and C
    without RSDs respectively (see Table \ref{table:EOR_param}). Red open circles show results
    for model A with RSDs from the MMRRM model. The measurements are derived from the
    $60$ largest triangles in our analysis, with an average scale of
    $\langle (r_1 r_2 r_3)^{1/3} \rangle = 73.8$ Mpc. The evolution of the
    linear and quadratic bias with neutral fraction is well fitted by the linear and quadratic
    polynomials from equation (\ref{eq:b1-nf_fit}) and (\ref{eq:b2-nf_fit}), respectively.
    Fits to model A without RSD are shown as red dashed-dotted lines).
}
\label{fig:b1b2_nf}
\end{figure}

The dependence of the our 3PCF bias measurements on the global neutral fraction
is displayed in Fig. \ref{fig:b1b2_nf} for all EoR models without RSD. In addition, for model
A we show results from simulations with MMRRM RSDs.
These measurements are derived from our largest scale bin with $60$ triangles
which have on average a scale of $\langle (r_1 r_2 r_3)^{1/3} \rangle = 73.8$ Mpc.
We find that the \nf{} dependence of the linear and quadratic bias from all models 
is well fitted by first and second-order polynomials respectively. Requiring that
$\delta_T$ is an unbiased tracer of the matter fluctuations at \nf{}$=1$
(i.e. $b_1=1$ and $b_2=0$) leads to the expressions
\begin{equation}
b_1 = \eta (\langle x_{\rm HI} \rangle - 1) + 1
\label{eq:b1-nf_fit}
\end{equation}
and
\begin{equation}
b_2 = \kappa \{ (\langle x_{\rm HI} \rangle - x_0)^2 - (1-x_0)^2 \}
\label{eq:b2-nf_fit}.
\end{equation}
Fitting these expressions to the results from model A without RSD
delivers $\eta = 14.80$, $\kappa = 272.30$ and $x_0=0.73$.
The bias measurements from the different models at fixed global neutral fractions
show an overall agreement of $\lesssim 20\%$, which lines up with the similarity
of the 3PCF measurements from different models, shown in the right panel of Fig. \ref{fig:3pc_EORmodels}.
It is interesting to note, that the bias from model A with RSD is close to results
derived without RSD, given the strong changes of the 21cm 3PCF amplitude
due to RSD, which we saw in Fig. \ref{fig:3pc_RSD}. This finding may be explained by the
non-linear dependence of the 21cm 3PCF on linear bias, shown in equation (\ref{eq:3pc_bias})
according to which small changes in the latter can cause strong changes in the former.

\begin{figure}
\centering\includegraphics[width=6.5 cm, angle=270]{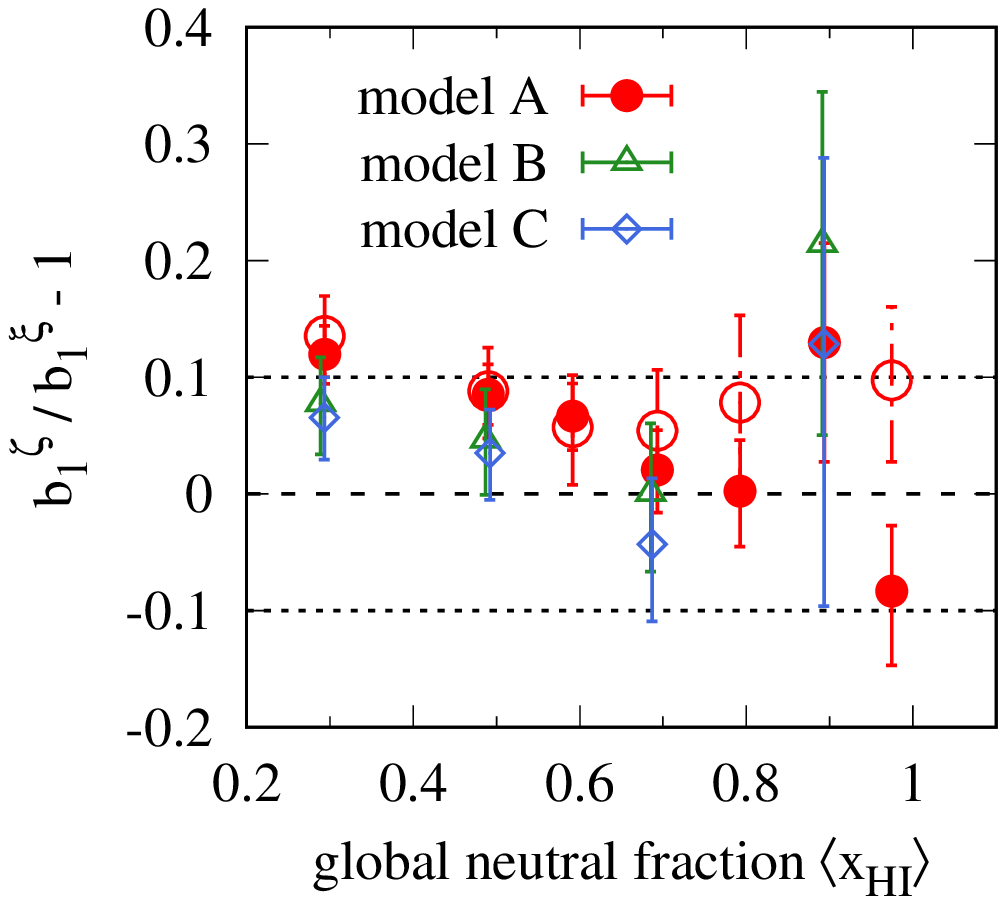}
\caption{Relative deviations between the linear bias measurements from the 21cm 3PCF,
shown in Fig. \ref{fig:b1b2_nf}, with respect to the corresponding measurements from
the 21cm 2PCF. Note that the result for model A with RSD (red open circle) at \nf$=0.9$
is off the chart with a value of $\sim 1.2$.}
\label{fig:b1_nf_reldiff}
\end{figure}

In Fig. \ref{fig:b1_nf_reldiff} we show the relative deviations of the different
$b_1$ measurements from Fig. \ref{fig:b1b2_nf} with respect to the corresponding
measurements, derived from fits to the 21cm 2PCF in the range of $40-100$ Mpc.
We thereby generalize the corresponding comparison from Fig. \ref{fig:b1b2_r}
to different EoR models and the case of model A with RSD.
We find the linear bias from both statistics to agree at the $\lesssim 10 \%$ level
in all considered cases, which indicates that the bias measurements are
physically meaningful. The fact that the $b_1$ measurements from the 2PCF and 3PCF
also agree when RSDs are included in the simulations further suggests that the quadratic
bias model may be useful for modeling observational data, albeit the meaning of
the bias parameters is not clear and will change in the presence of RSD.

\subsection{Direct measurements of the $\delta_m-\delta_{\delta T}$ bias relation}\label{sec:scatter_plot}

\begin{figure*}
\centering\includegraphics[width=11 cm, angle=270]{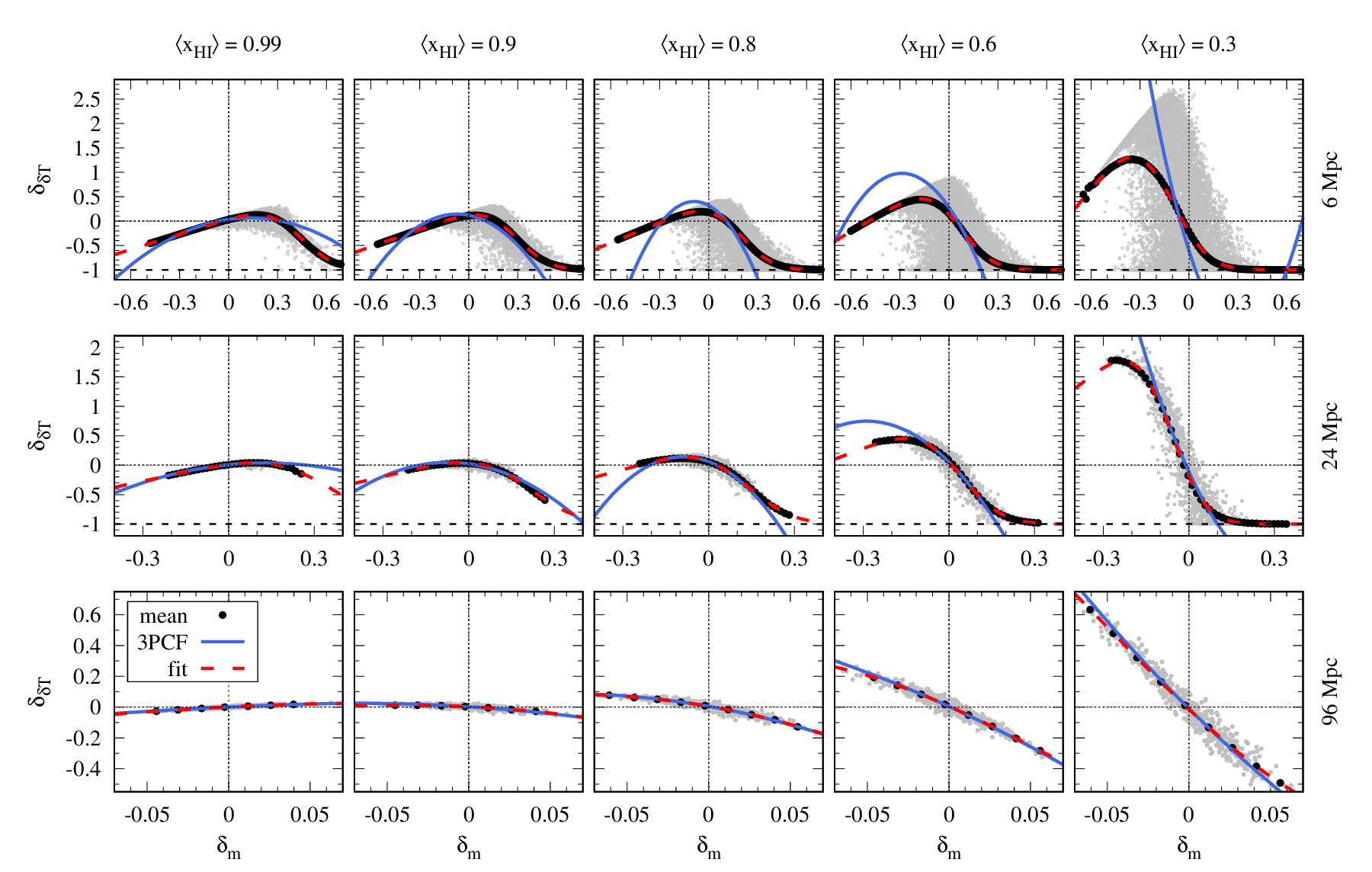}
\caption{Spatial fluctuations of the 21cm brightness temperature \dT{} versus those in the matter density.
Grey dots show direct measurements from one realization, while each dot represents a measurement
in one grid cell. Black dots show the mean \dT{} fluctuations in bins of $\delta_m$, averaged
over $200$ realizations. Red dashed lines show fits to equation (\ref{eq:21cm_biasmodel}). Blue solid lines
show the quadratic bias model from equation (\ref{eq:quad_bias_model}) with the bias parameters
measured from the 3PCF, derived from the $60$ largest triangles in our analysis, with
$\langle (r_1 r_2 r_3)^{1/3} \rangle \simeq 73.8$ Mpc, shown in Fig. \ref{fig:b1b2_r}.
Columns show results for different volume weighted global neutral fractions \nf{}. The top, central and bottom rows
show results for grid cells with side lengths of $6$, $24$ and $96$ Mpc, respectively. Direct measurements
are diluted for each cell size differently for clarity. Note that the 3PCF parameters for the quadratic bias model
are the same for different cell sizes, but vary for different \nf{}.}
\label{fig:dmdT-scatter}
\end{figure*}

We now aim at a more direct test of the quadratic bias model, taking advantage of the
fact that in the simulations we have access to both, the spatial fluctuations in the
21cm brightness temperature as well as those in the matter density.  Having these quantities defined
in smoothing volumes (i.e. grid cells) at different positions $\bf x$, enables us to directly
measure the bias relation $\delta_m - \delta_{\delta T}$
and verify how well it can be described by a deterministic function
$\delta_{\delta T}({\bf x}) = F[\delta_m({\bf x})]$. Our particular interest lies thereby
on testing how well this relation agrees with the  quadratic bias model from
equation (\ref{eq:quad_bias_model}), when using the bias parameters from the 3PCF measurements.

We show in Fig. \ref{fig:dmdT-scatter} a scatter plot of the $\delta_m - \delta_{\delta T}$ relation,
for one realization of model A without RSD as grey dots, while each dot represents $\delta_m$ and $\delta_{\delta T}$
measurements from one grid cell of the simulation. Results are shown for cells with side lengths of $6$, $24$
and $96$ Mpc at redshifts with neutral fractions \nf\ between $0.99$ and $0.6$.

The results for $6$ Mpc grid cells show that, at very early times (\nf$= 0.99$ ),
the 21cm brightness temperature fluctuations $\delta_{\delta T}$ in most cells follow a
roughly linear relation with respect to the matter density fluctuations, as expected from
equation (\ref{eq:dT}). A departure from this relation
occurs for cells with high matter overdensities, where the 21cm brightness temperature
is lower due to a higher fraction of ionized hydrogen. The effect becomes stronger at lower
redshifts, as the reionization proceeds. At $z=10.34$ with \nf$=0.8$, 
we already find a significant fraction of completely ionized cells with $\delta_{\delta T}=-1$.
The trend is apparent for all cell sizes, and provides an understanding
of why the linear bias from the correlation function measurements changes its sign
from positive to negative with decreasing redshift, as discussed in Section \ref{sec:bias_2PCFvs3PCF}.
The linear bias corresponds to the slope of the $\delta_m - \delta_{\delta T}$ relation
at $\delta_m=0$. The results in Fig. \ref{fig:dmdT-scatter} show clearly how
this slope follows the trend expected from our previous $b_1$ measurements, while its
value depends slightly on the cell size. For a more detailed comparison with the quadratic bias model
predictions, we show in Fig. \ref{fig:dmdT-scatter} the mean $\delta_m - \delta_{\delta T}$
relation in bins of $\delta_m$ as black dots. We find this mean to be well approximated by fits to 
\begin{equation}
	\delta_{\delta T} = \alpha \erfc \left( \frac{\delta_m-\beta}{\gamma}\right)(1+\delta_m)-1
	\label{eq:21cm_biasmodel}
\end{equation}
at all considered redshifts and scales (red dashed lines in Fig. \ref{fig:dmdT-scatter}),
where $\alpha$, $\beta$ and $\gamma$ are free parameters.
In addition to these fits, we show the prediction of the quadratic
bias model with 3PCF bias parameters as solid blue lines. The 3PCF bias parameters
were measured, using $60$ triangles with $\langle (r_1r_2r_3)^{1/3} \rangle \simeq 73.8$ Mpc,
covering the largest scales in our triangles sample (see Fig. \ref{fig:b1b2_r}). Comparing the
quadratic bias model to the mean from the direct measurements in Fig. \ref{fig:dmdT-scatter},
we find a good agreement for large grid cells with $96$ Mpc side lengths.
For cells with side $24$ Mpc side lengths, the quadratic model is a good approximation at 
$\delta_m \simeq 0$, while clear deviations occur at the tails of the $\delta_m$ distribution.
For the grid of $6$ Mpc cells, on which we measured most of the correlation functions
(see Section \ref{sec:3pc}), we find that the quadratic bias model differs strongly from the direct
measurements for all values of $\delta_m$. This result indicates that the residuals
from the mean bias relation may not be random, but spatially correlated. They can hence affect
the $\delta_{\delta T}$ values at larger smoothing scales, and modify the
measured bias parameters from the 3PCF fits. The agreement between the 3PCF bias prediction
and the $\delta_m - \delta_{\delta T}$ relation at large scales indicates, that both statistics
are affected by the residuals in the same way, which is subject of our ongoing investigations.

\begin{figure}
\centering\includegraphics[width=7 cm, angle=270]{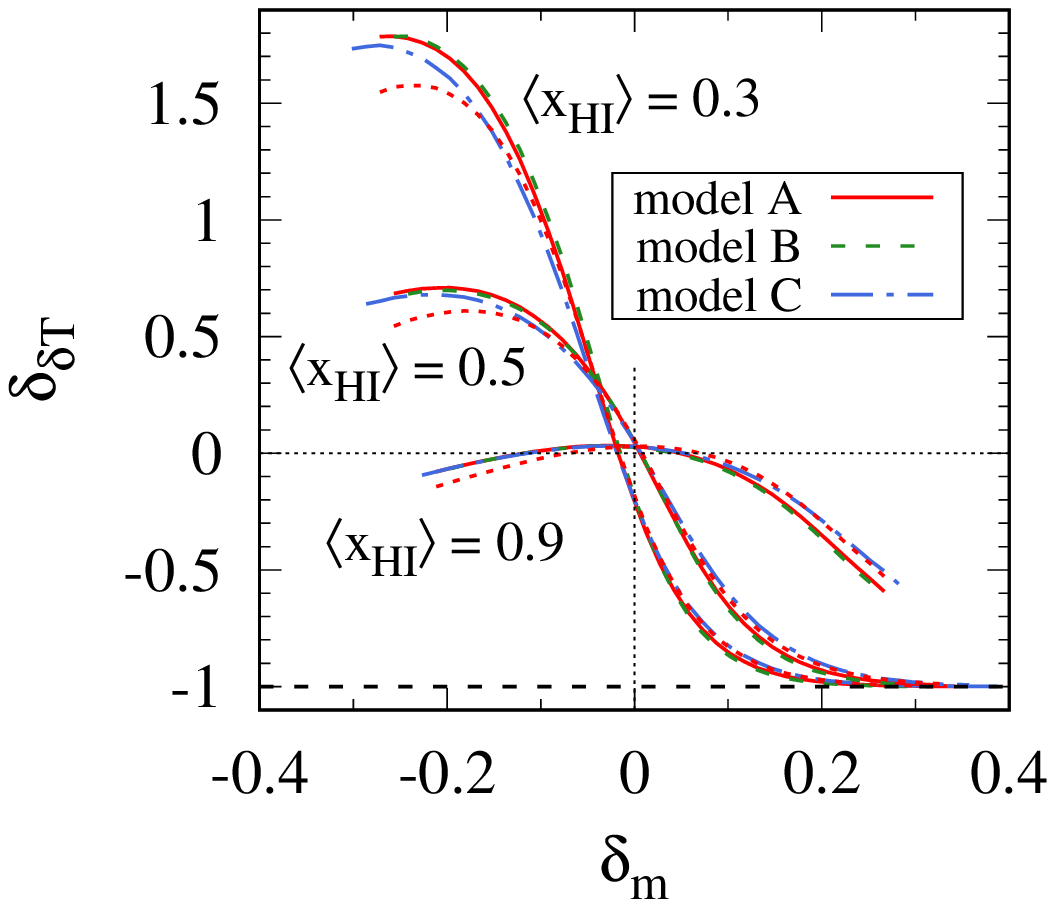}
\caption{Mean fluctuations in the 21cm brightness temperature \dT{} in bins of the matter overdensity $\delta_m$ for different EoR models at different global volume weighted neutral fractions \nf.
Results for models A, B and C (as indicated in the legend) are based on simulations without RSD,
while we show for model A also results with RSD from the MMRRM model as doted lines.}
\label{fig:dmdT-mean_models}
\end{figure}

In the previous Sections \ref{sec:bias_comparison}, we found that the 21cm 2PCF and 3PCF,
as well as the bias parameters are almost independent of the EoR model at fixed volume weighted
global neutral fractions \nf\ (i.e. Fig. \ref{fig:2pc_EORmodels}, \ref{fig:3pc_EORmodels} and
\ref{fig:b1b2_nf}), which suggests, that also the $\delta_m - \delta_{\delta T}$ relation
should only depend weakly on the EoR model at fixed \nf. We test if this assumption holds,
by comparing the mean $\delta_m - \delta_{\delta T}$ relation for model A, B and C without RSD
in Fig. \ref{fig:dmdT-mean_models} at \nf\ values between $0.9$ and $0.3$ for a smoothing scale
of $24$ Mpc. Our results line up with our previous findings, as we find very similar results
from different EoR models. This implies, that the different 21cm statistics are almost entirely
determined by the global neutral fraction. However, this result may be specific to the excursion set
modeling, used by \texttt{21cmFAST}. In addition to results without RSD, we show in
Fig. \ref{fig:dmdT-mean_models} results with RSD from model A as red dotted lines.
We find that at early times, when \nf$=0.9$, RSDs enhance (decrease) the 21cm fluctuations in underdense
(overdense) regions, while at later times the RSD effect in underdense regions is dominant.
The latter finding may be understood from the fact that at later times less neutral gas in present
overdense regions, which decreases the overall effect of RSDs on $\delta_{dT}$ for high $\delta_m$.
As a result, the fraction of regions, which are affected by RSDs decreases with time, which lines
up with the lower impact of RSDs on the 21cm correlation functions at late times, which we found in
Fig. \ref{fig:2pc_RSD} and \ref{fig:3pc_RSD}. However, a conclusive interpretation of these results
requires further investigations.

\section{Summary and Conclusions}\label{sec:conclusion}


We studied 2PCF and 3PCF of the 21cm brightness temperature  during the epoch of reionization.
The goal of our study was to characterize the 21cm 2PCF and, for the first time, the 21cm 3PCF
in configuration space for different redshifts, scales and triangle shapes, using measurements in simulations.
Based on these measurements we tested how well these 21cm correlation functions can be
described by the local quadratic bias model, which has been commonly employed for relating the
2PCF and 3PCF of galaxies and halos to the corresponding statistics of the underlying dark
matter density field.

Our simulations were produced by the semi-numerical code 21cmFAST \citep{Mesinger11},
using three different combinations of EoR model parameters, to which we refer to as model A, B and C, as
summarized in Table \ref{table:EOR_param}. For each parameter combination we generated
$100-200$ realizations of the 21cm brightness
temperature and the underlying matter density field with different random initial conditions.
Each set of simulations covers a total volume of up to $\simeq (4.5 \ \text{Gpc})^3$, providing small
errors as well as error covariance estimates for our 2PCF and 3PCF measurements, which allows for a detailed
comparison with the bias model predictions.
Redshift space distortions (RSDs) are neglected in the main part of our analysis, which we discuss in the following,
to simplify the interpretation of our results. However, for our fiducial model A, we investigated how the 21cm correlations
and the corresponding bias measurements are affected by RSDs, which is briefly discussed towards the end of this section.

Our 21cm 2PCF measurements present a strong redshift evolution, with an amplitude change of $2-3$ orders
of magnitude for neutral fractions in the range 
$0.3 \lesssim \langle x_{\rm HI} \rangle \lesssim 0.99$ ($8.4 \lesssim z \lesssim 13.5$)
for model A, which demonstrates the high sensitivity of the 21cm 2PCF
to the state of reionization. This redshift evolution is also highly sensitive to the EoR model parameters,
while results from different parameter combinations show a percent level agreement at all considered scales
when compared at fixed global neutral fractions (Fig. \ref{fig:2pc_EORmodels}).
The  quadratic expansion of the local bias model in equation (\ref{eq:quad_bias_model}) predicts that the
matter and 21cm 2PCF are related to each other by the scale independent linear bias factor $b_1$
at leading order. We show in Fig. \ref{fig:2pc_b1} that this prediction can describe the measurements
at scales between $30$ and $100$ Mpc with $\lesssim10\%$ ($\lesssim30\%$) accuracy for \nf$\gtrsim0.7$ ($\gtrsim0.3$).
This means that the impact of the reionization on the large-scale 21cm 2PCF can be well characterized by a single parameter,
which limits the constraining power of this statistics for reionization models.
Strong deviations from the leading order 2PCF bias model occur at scales of $\lesssim 30$ Mpc, which corresponds
roughly to the typical size of ionized regions in our simulation (Fig. \ref{fig:bubble_sizes}). This finding
indicates that patchy reionization needs to be taken into account for a detailed modeling
of the 21cm 2PCF at small scales and late times, as it has been discussed in the literature \citep[e.g.][]{FZH04}.

Our measurements of the 21cm 3PCF show a strong dependence on the triangle opening angle,
as well as on the overall length scale defined as $(r_1 r_2 r_3)^{1/3}$ (see e.g. Fig. \ref{fig:3pc_allconfigs}).
For a fixed triangle shape, we find a stronger change of the amplitude than for the 2PCF
by up to four orders of magnitude in the considered redshift range (Fig. \ref{fig:3pc_z}).
The dependence on the opening angle shows that the 3PCF does not only probe the
scale dependence of the 21cm fluctuations, but also their morphology. It therefore provides
access to additional information in the 21cm signal besides its non-Gaussianity, to which the 2PCF is not sensitive.
As for the 2PCF, we find a strong dependence of the 21cm 3PCF on the EoR model parameters
at fixed redshifts, while results from different parameter combinations agree at the $\simeq 10\%$ level
(Fig. \ref{fig:3pc_EORmodels}) at fixed global neutral fractions.

The leading-order approximation of 21cm 3PCF, based on the bias model from equation (\ref{eq:3pc_bias}),
delivers fits which are in $\sim 2\sigma$ agreement with our measurements (Fig. \ref{fig:3pc_allconfigs})
for triangles with $(r_1r_2r_3)^{1/3} \gtrsim 60$ Mpc. This result indicates a good performance of the
bias model, given our small measurements errors. A notable small scale feature of our
3PCF measurements, which is not described by the bias model, is the increase of the amplitude at the smallest
scales at early times (\nf{}$\gtrsim 0.9$, Fig. \ref{fig:3pc_z}). Such an increase has also been reported by \citet{Majumdar17} for bispectrum measurements in simulations, as well model predictions, based on randomly
distributed ionized bubbles.

The dependence of our best fit bias parameters on the chosen triangle scale range, shown in Fig. \ref{fig:b1b2_r},
is within the expected $1-2 \sigma$ uncertainty for \nf$\gtrsim 0.7$. At \nf$=0.6$ this scale dependence becomes
more significant, indicating a breakdown of the leading order 3PCF bias model at late times of reioinization, which
we already noticed in the 2PCF analysis (Fig. \ref{fig:2pc_b1}).
%
The linear bias from the 3PCF is compared to the results from the 2PCF in Fig. \ref{fig:b1_nf_reldiff}.
The comparison reveals a $\simeq 10\%$ agreement for all EoR parameter combinations in the considered redshift
range (with \nf$>0.3$), when the analysis is restricted to large triangles with $(r_1r_2r_3)^{1/3}\gtrsim60$ Mpc.
This result is consistent with the aforementioned inaccuracy of the linear bias model for the 2PCF at large scales.

Our comparison with results from approximate N-body simulations based on the COLA method, shown in
Fig. \ref{fig:2pc_EORmodels}, \ref{fig:3pc_configs} and \ref{fig:3pc_z}, revealed that our main results
for the correlations of the matter and 21cm fields are only weakly affected by the usage of the Zel'dovich
approximation at large scales. We therefore consider our conclusions drawn from these measurements to be
robust in this regard. Note that this would not be the case at small scales and low redshifts \citep[][]{Scoccimarro97, Leclercq13}.
The good agreement between the linear bias from the two- and three-point statistics further indicates that the
quadratic bias model is physically meaningful, in particular at large scales and early times. The simple polynomial relations
between the linear and quadratic bias with respect to the global neutral fraction, shown in Fig. \ref{fig:b1b2_nf},
further indicate that there may be a simple physical interpretation of the bias parameters \citep[see also][]{mqda18}.

Up to this point we tested the quadratic bias model for the assumed deterministic $\delta_{\delta T} = F(\delta_m)$
relation via its leading order prediction for the 2PCF and 3PCF. In the final step of our analysis, we
conducted a more direct test of the model, by comparing its predicted $\delta_m - \delta_{\delta T}$ relation to direct
measurements of this relation in the simulations (Fig. \ref{fig:dmdT-scatter}). For this comparison, we employ the
linear and quadratic bias parameters, measured from the 3PCF at large scales.  We find strong deviations
between the bias model prediction and the direct average value of $\delta_{\delta T}$ in bins
of $\delta_m$ when using small smoothing scales of $6$ Mpc for the direct measurements. However, for large
smoothing scales of $96$ Mpc, we find very good agreement in the considered redshift range.
Another interesting finding is that the average $\delta_m - \delta_{\delta T}$ relation is well approximated
by the analytic expression, given by equation (\ref{eq:21cm_biasmodel}). The physical interpretation of
this expression may provide the aforementioned interpretation of the bias parameters and is the subject of our followup work.

The results discussed above refer to our simulations without RSDs. We included the latter
to our simulations with EoR model A using a simplistic RSD model, which was implemented in 21cmFAST.
We compared these results to those based on the physically better motivated MMRRM model for RSDs from \citet{Mao12}.
For both RSD models we find a stronger impact of RSDs on the 21cm 2PCF at early times of reionization
(Fig. \ref{fig:2pc_RSD}), confirming results from the literature for the 21cm power spectrum
\citep[e.g.][]{McQuinn06, MesingerFurlanetto07}. We find the same trend for the 21cm 3PCF, while
in that case the change of the amplitude is overall stronger than for the 2PCF and highly dependent on
the scale, as well as on the triangle opening angle (Fig. \ref{fig:3pc_RSD}). For both statistics
the impact of RSDs from the MMRRM model are significantly stronger than predicted by the simplistic model
from 21cmFAST, in particular at late times with low global neutral fractions. It is interesting to note that
the bias parameters are only weakly affected by RSDs, despite of the strong change in of the correlations functions.
This finding may result from the non-linear relation between the linear bias and correlation functions,
given by equation (\ref{eq:2pc_bias}) and (\ref{eq:3pc_bias}).

Overall our results show that the quadratic bias model provides a consistent description of the 21cm 2PCF and 3PCF
in \texttt{21cmFAST} simulations on scales larger than $\simeq 30$ Mpc at early times of the reionization
with \nf$> 0.6$. It further describes the $\delta_m - \delta_{\delta T}$ well at large smoothing scales of $\gtrsim 30$ Mpc. The latter finding suggests that the linear and quadratic bias parameter measurements from the 21cm 3PCF can provide
insights to the process leading to the reionization from upcoming 21cm observations.
In principle, the linear and quadratic bias parameters, $b_1$ and $b_2$, depend on the underlying model that 
predicts both global reionization history and the clustering and morphology of ionized regions. 
Measurements of the 21cm 3PCF in upcoming radio observations, therefore, can improve constraints 
on astrophysical processes driving reionization upon those from the 2PCF.
We further find that a detailed interpretation of the bias parameters from the observed 21cm 3PCF
in three dimension 21cm mapping further requires an understanding of how the 21cm 3PCF is affected by
RSDs.
A general limitation for our approach is given by the fact that our analysis is based on fluctuations of the brightness
temperature around the global mean, i.e.\ the monopole of the 21cm signal. Nevertheless, the 21cm mean signal cannot be
observed with radio interferometers such as SKA and HERA, and would therefore need to be obtained from single antenna experiments.

Recently \citet{Beane18} found that the quadratic bias model also provides reasonable fits to measurements of
the three-point cross-bispectrum between maps of the 21cm brightness temperature and the CII emission line
intensity in simulations. This work showed on one hand, that the bias model still holds
when being applied on the 21cm signal, instead of its fluctuations. This can be expected, given that
the underlying assumption of a deterministic bias function is not restricted to a given definition
of the observed quantity, as long as deviations around the expansion point are sufficiently small on average.
On the other hand, their work demonstrated the applicability of the bias modeling in Fourier space,
which is more closely related to future observational data sets. However, the configuration space
analysis, on which the present work is based, simplifies the physical interpretation of the results,
such as the aforementioned impact of residuals around the mean $\delta_{\delta T} - \delta_m$ bias relation on the clustering.
The latter may be important to understand shortcomings of the bias model revealed in this work, in particular
at small scales and late times of reionization. An improvement of the bias model might require further investigation
of such residuals as well as an expansion of the 21cm 2PCF and 3PCF beyond the leading order. Such improvements
were recently investigated by \citet{mqda18}, who found that adding a dependence on the wave number
and an effective bubble size parameter to the quadratic terms in the bias expansion provides better fits to the
21cm power spectrum at late times of reionization.


\section*{Acknowledgments}
YM and HJM were supported in part by the National Key Basic Research and Development Program of China
(No.2018YFA0404502, No.2018YFA0404503),  and by the NSFC-ISF joint research program (NSFC Grant No.11761141012),
and the National Natural Science Foundation of China (``NSFC'', Grant No.11821303). 
YM was also supported in part by the NSFC (Grant No.11673014, 11543006), 
by the National Key Program for Science and Technology Research and Development of China (No.2017YFB0203302), 
by the Chinese National Thousand Youth Talents Program, and by the Opening Project of 
Key Laboratory of Computational Astrophysics, National Astronomical Observatories, Chinese Academy of Sciences. 
KH was supported in part by the NSFC (Grant No.11750110419), 
by the China Postdoctoral Science Foundation, and the International Postdoctoral Fellowship from the 
Ministry of Education and the State Administration of Foreign Experts Affairs of China.  HJM was also
supported in part by the NSFC (Grant No.11673015,11733004,11761131004) and NSF AST-1517528.
BDW acknowledges support from the Simons Foundation.

\bibliographystyle{mnras}
\bibliography{references.bib}

\providecommand{\noopsort}[1]{}\providecommand{\singleletter}[1]{#1}%
\begin{thebibliography}{}
\makeatletter
\relax
\def\mn@urlcharsother{\let\do\@makeother \do\$\do\&\do\#\do\^\do\_\do\%\do\~}
\def\mn@doi{\begingroup\mn@urlcharsother \@ifnextchar [ {\mn@doi@}
  {\mn@doi@[]}}
\def\mn@doi@[#1]#2{\def\@tempa{#1}\ifx\@tempa\@empty \href
  {http://dx.doi.org/#2} {doi:#2}\else \href {http://dx.doi.org/#2} {#1}\fi
  \endgroup}
\def\mn@eprint#1#2{\mn@eprint@#1:#2::\@nil}
\def\mn@eprint@arXiv#1{\href {http://arxiv.org/abs/#1} {{\tt arXiv:#1}}}
\def\mn@eprint@dblp#1{\href {http://dblp.uni-trier.de/rec/bibtex/#1.xml}
  {dblp:#1}}
\def\mn@eprint@#1:#2:#3:#4\@nil{\def\@tempa {#1}\def\@tempb {#2}\def\@tempc
  {#3}\ifx \@tempc \@empty \let \@tempc \@tempb \let \@tempb \@tempa \fi \ifx
  \@tempb \@empty \def\@tempb {arXiv}\fi \@ifundefined
  {mn@eprint@\@tempb}{\@tempb:\@tempc}{\expandafter \expandafter \csname
  mn@eprint@\@tempb\endcsname \expandafter{\@tempc}}}

\bibitem[\protect\citeauthoryear{{Baldauf}, {Seljak}, {Desjacques}  \&
  {McDonald}}{{Baldauf} et~al.}{2012}]{Baldauf12}
{Baldauf} T.,  {Seljak} U.,  {Desjacques} V.,   {McDonald} P.,  2012, \mn@doi
  [\prd] {10.1103/PhysRevD.86.083540}, \href
  {http://adsabs.harvard.edu/abs/2012PhRvD..86h3540B} {86, 083540}

\bibitem[\protect\citeauthoryear{{Barkana} \& {Loeb}}{{Barkana} \&
  {Loeb}}{2001}]{BarkanaLoeb01}
{Barkana} R.,  {Loeb} A.,  2001, \mn@doi [\physrep]
  {10.1016/S0370-1573(01)00019-9}, \href
  {http://adsabs.harvard.edu/abs/2001PhR...349..125B} {349, 125}

\bibitem[\protect\citeauthoryear{{Barriga} \& {Gazta{\~n}aga}}{{Barriga} \&
  {Gazta{\~n}aga}}{2002}]{BaGaz02}
{Barriga} J.,  {Gazta{\~n}aga} E.,  2002, \mn@doi [\mnras]
  {10.1046/j.1365-8711.2002.05431.x}, \href
  {http://adsabs.harvard.edu/abs/2002MNRAS.333..443B} {333, 443}

\bibitem[\protect\citeauthoryear{{Beane} \& {Lidz}}{{Beane} \&
  {Lidz}}{2018}]{Beane18}
{Beane} A.,  {Lidz} A.,  2018, \mn@doi [\apj] {10.3847/1538-4357/aae388}, \href
  {https://ui.adsabs.harvard.edu/abs/2018ApJ...867...26B} {867, 26}

\bibitem[\protect\citeauthoryear{{Bernardeau}, {Colombi}, {Gazta{\~n}aga}  \&
  {Scoccimarro}}{{Bernardeau} et~al.}{2002}]{Bernardeau02}
{Bernardeau} F.,  {Colombi} S.,  {Gazta{\~n}aga} E.,   {Scoccimarro} R.,  2002,
  \mn@doi [\physrep] {10.1016/S0370-1573(02)00135-7}, \href
  {http://adsabs.harvard.edu/abs/2002PhR...367....1B} {367, 1}

\bibitem[\protect\citeauthoryear{{Bharadwaj} \& {Pandey}}{{Bharadwaj} \&
  {Pandey}}{2005}]{Bharadwaj05}
{Bharadwaj} S.,  {Pandey} S.~K.,  2005, \mn@doi [\mnras]
  {10.1111/j.1365-2966.2005.08836.x}, \href
  {http://adsabs.harvard.edu/abs/2005MNRAS.358..968B} {358, 968}

\bibitem[\protect\citeauthoryear{{Bryan} \& {Norman}}{{Bryan} \&
  {Norman}}{1998}]{BryNo98}
{Bryan} G.~L.,  {Norman} M.~L.,  1998, \mn@doi [\apj] {10.1086/305262}, \href
  {https://ui.adsabs.harvard.edu/#abs/1998ApJ...495...80B} {495, 80}

\bibitem[\protect\citeauthoryear{{Chan}, {Scoccimarro}  \& {Sheth}}{{Chan}
  et~al.}{2012}]{css12}
{Chan} K.~C.,  {Scoccimarro} R.,   {Sheth} R.~K.,  2012, \mn@doi [\prd]
  {10.1103/PhysRevD.85.083509}, \href
  {http://adsabs.harvard.edu/abs/2012PhRvD..85h3509C} {85, 083509}

\bibitem[\protect\citeauthoryear{{Eisenstein} \& {Hu}}{{Eisenstein} \&
  {Hu}}{1998}]{EH98}
{Eisenstein} D.~J.,  {Hu} W.,  1998, \mn@doi [\apj] {10.1086/305424}, \href
  {http://adsabs.harvard.edu/abs/1998ApJ...496..605E} {496, 605}

\bibitem[\protect\citeauthoryear{{Field}}{{Field}}{1958}]{Field58}
{Field} G.~B.,  1958, \mn@doi [Proceedings of the IRE]
  {10.1109/JRPROC.1958.286741}, \href
  {http://adsabs.harvard.edu/abs/1958PIRE...46..240F} {46, 240}

\bibitem[\protect\citeauthoryear{{Fry} \& {Gaztanaga}}{{Fry} \&
  {Gaztanaga}}{1993}]{FryGa93}
{Fry} J.~N.,  {Gaztanaga} E.,  1993, \mn@doi [\apj] {10.1086/173015}, \href
  {http://adsabs.harvard.edu/abs/1993ApJ...413..447F} {413, 447}

\bibitem[\protect\citeauthoryear{{Furlanetto}}{{Furlanetto}}{2006}]{Furlanetto06}
{Furlanetto} S.~R.,  2006, \mn@doi [\mnras] {10.1111/j.1365-2966.2006.10725.x},
  \href {http://adsabs.harvard.edu/abs/2006MNRAS.371..867F} {371, 867}

\bibitem[\protect\citeauthoryear{{Furlanetto}, {Zaldarriaga}  \&
  {Hernquist}}{{Furlanetto} et~al.}{2004}]{FZH04}
{Furlanetto} S.~R.,  {Zaldarriaga} M.,   {Hernquist} L.,  2004, \mn@doi [\apj]
  {10.1086/423025}, \href {http://adsabs.harvard.edu/abs/2004ApJ...613....1F}
  {613, 1}

\bibitem[\protect\citeauthoryear{{Furlanetto}, {Oh}  \& {Briggs}}{{Furlanetto}
  et~al.}{2006}]{FOB06}
{Furlanetto} S.~R.,  {Oh} S.~P.,   {Briggs} F.~H.,  2006, \mn@doi [\physrep]
  {10.1016/j.physrep.2006.08.002}, \href
  {http://adsabs.harvard.edu/abs/2006PhR...433..181F} {433, 181}

\bibitem[\protect\citeauthoryear{{Gazta{\~n}aga} \&
  {Scoccimarro}}{{Gazta{\~n}aga} \& {Scoccimarro}}{2005}]{GazSco05}
{Gazta{\~n}aga} E.,  {Scoccimarro} R.,  2005, \mn@doi [\mnras]
  {10.1111/j.1365-2966.2005.09234.x}, \href
  {http://adsabs.harvard.edu/abs/2005MNRAS.361..824G} {361, 824}

\bibitem[\protect\citeauthoryear{{Hartlap}, {Simon}  \& {Schneider}}{{Hartlap}
  et~al.}{2007}]{Hartlap07}
{Hartlap} J.,  {Simon} P.,   {Schneider} P.,  2007, \mn@doi [\aap]
  {10.1051/0004-6361:20066170}, \href
  {http://adsabs.harvard.edu/abs/2007A%26A...464..399H} {464, 399}

\bibitem[\protect\citeauthoryear{{Hoffmann}, {Gazta{\~n}aga}, {Scoccimarro}  \&
  {Crocce}}{{Hoffmann} et~al.}{2018}]{Hoffmann17}
{Hoffmann} K.,  {Gazta{\~n}aga} E.,  {Scoccimarro} R.,   {Crocce} M.,  2018,
  \mn@doi [\mnras] {10.1093/mnras/sty187}, \href
  {https://ui.adsabs.harvard.edu/abs/2018MNRAS.476..814H} {476, 814}

\bibitem[\protect\citeauthoryear{{Howlett}, {Manera}  \& {Percival}}{{Howlett}
  et~al.}{2015}]{Howlett15}
{Howlett} C.,  {Manera} M.,   {Percival} W.~J.,  2015, \mn@doi [Astronomy and
  Computing] {10.1016/j.ascom.2015.07.003}, \href
  {https://ui.adsabs.harvard.edu/abs/2015A&C....12..109H} {12, 109}

\bibitem[\protect\citeauthoryear{{Leclercq}, {Jasche}, {Gil-Mar{\'{\i}}n}  \&
  {Wandelt}}{{Leclercq} et~al.}{2013}]{Leclercq13}
{Leclercq} F.,  {Jasche} J.,  {Gil-Mar{\'{\i}}n} H.,   {Wandelt} B.,  2013,
  \mn@doi [\jcap] {10.1088/1475-7516/2013/11/048}, \href
  {http://adsabs.harvard.edu/abs/2013JCAP...11..048L} {11, 048}

\bibitem[\protect\citeauthoryear{{Lidz}, {Zahn}, {McQuinn}, {Zaldarriaga},
  {Dutta}  \& {Hernquist}}{{Lidz} et~al.}{2007}]{Lidz07}
{Lidz} A.,  {Zahn} O.,  {McQuinn} M.,  {Zaldarriaga} M.,  {Dutta} S.,
  {Hernquist} L.,  2007, \mn@doi [\apj] {10.1086/511670}, \href
  {http://adsabs.harvard.edu/abs/2007ApJ...659..865L} {659, 865}

\bibitem[\protect\citeauthoryear{{Liu} \& {Parsons}}{{Liu} \&
  {Parsons}}{2016}]{Liu16}
{Liu} A.,  {Parsons} A.~R.,  2016, \mn@doi [\mnras] {10.1093/mnras/stw071},
  \href {http://adsabs.harvard.edu/abs/2016MNRAS.457.1864L} {457, 1864}

\bibitem[\protect\citeauthoryear{{Majumdar}, {Pritchard}, {Mondal},
  {Watkinson}, {Bharadwaj}  \& {Mellema}}{{Majumdar} et~al.}{2018}]{Majumdar17}
{Majumdar} S.,  {Pritchard} J.~R.,  {Mondal} R.,  {Watkinson} C.~A.,
  {Bharadwaj} S.,   {Mellema} G.,  2018, \mn@doi [\mnras]
  {10.1093/mnras/sty535}, \href
  {https://ui.adsabs.harvard.edu/abs/2018MNRAS.476.4007M} {476, 4007}

\bibitem[\protect\citeauthoryear{{Mao}, {Shapiro}, {Mellema}, {Iliev}, {Koda}
  \& {Ahn}}{{Mao} et~al.}{2012}]{Mao12}
{Mao} Y.,  {Shapiro} P.~R.,  {Mellema} G.,  {Iliev} I.~T.,  {Koda} J.,   {Ahn}
  K.,  2012, \mn@doi [\mnras] {10.1111/j.1365-2966.2012.20471.x}, \href
  {http://adsabs.harvard.edu/abs/2012MNRAS.422..926M} {422, 926}

\bibitem[\protect\citeauthoryear{{McQuinn} \& {D'Aloisio}}{{McQuinn} \&
  {D'Aloisio}}{2018}]{mqda18}
{McQuinn} M.,  {D'Aloisio} A.,  2018, \mn@doi [Journal of Cosmology and
  Astro-Particle Physics] {10.1088/1475-7516/2018/10/016}, \href
  {https://ui.adsabs.harvard.edu/abs/2018JCAP...10..016M} {2018, 016}

\bibitem[\protect\citeauthoryear{{McQuinn}, {Zahn}, {Zaldarriaga}, {Hernquist}
  \& {Furlanetto}}{{McQuinn} et~al.}{2006}]{McQuinn06}
{McQuinn} M.,  {Zahn} O.,  {Zaldarriaga} M.,  {Hernquist} L.,   {Furlanetto}
  S.~R.,  2006, \mn@doi [\apj] {10.1086/505167}, \href
  {http://adsabs.harvard.edu/abs/2006ApJ...653..815M} {653, 815}

\bibitem[\protect\citeauthoryear{{Mesinger} \& {Furlanetto}}{{Mesinger} \&
  {Furlanetto}}{2007}]{MesingerFurlanetto07}
{Mesinger} A.,  {Furlanetto} S.,  2007, \mn@doi [\apj] {10.1086/521806}, \href
  {http://adsabs.harvard.edu/abs/2007ApJ...669..663M} {669, 663}

\bibitem[\protect\citeauthoryear{{Mesinger}, {Furlanetto}  \& {Cen}}{{Mesinger}
  et~al.}{2011}]{Mesinger11}
{Mesinger} A.,  {Furlanetto} S.,   {Cen} R.,  2011, \mn@doi [\mnras]
  {10.1111/j.1365-2966.2010.17731.x}, \href
  {http://adsabs.harvard.edu/abs/2011MNRAS.411..955M} {411, 955}

\bibitem[\protect\citeauthoryear{{Pober} et~al.,}{{Pober}
  et~al.}{2014}]{Pober14}
{Pober} J.~C.,  et~al., 2014, \mn@doi [\apj] {10.1088/0004-637X/782/2/66},
  \href {http://adsabs.harvard.edu/abs/2014ApJ...782...66P} {782, 66}

\bibitem[\protect\citeauthoryear{{Pritchard} \& {Furlanetto}}{{Pritchard} \&
  {Furlanetto}}{2007}]{Pritchard06}
{Pritchard} J.~R.,  {Furlanetto} S.~R.,  2007, \mn@doi [\mnras]
  {10.1111/j.1365-2966.2007.11519.x}, \href
  {http://adsabs.harvard.edu/abs/2007MNRAS.376.1680P} {376, 1680}

\bibitem[\protect\citeauthoryear{{Pritchard} \& {Loeb}}{{Pritchard} \&
  {Loeb}}{2012}]{PL12}
{Pritchard} J.~R.,  {Loeb} A.,  2012, \mn@doi [Reports on Progress in Physics]
  {10.1088/0034-4885/75/8/086901}, \href
  {http://adsabs.harvard.edu/abs/2012RPPh...75h6901P} {75, 086901}

\bibitem[\protect\citeauthoryear{{Raste} \& {Sethi}}{{Raste} \&
  {Sethi}}{2018}]{Raste17}
{Raste} J.,  {Sethi} S.,  2018, \mn@doi [\apj] {10.3847/1538-4357/aac2d8},
  \href {https://ui.adsabs.harvard.edu/abs/2018ApJ...860...55R} {860, 55}

\bibitem[\protect\citeauthoryear{{Schmit} \& {Pritchard}}{{Schmit} \&
  {Pritchard}}{2018}]{SchmitPritchard17}
{Schmit} C.~J.,  {Pritchard} J.~R.,  2018, \mn@doi [\mnras]
  {10.1093/mnras/stx3292}, \href
  {https://ui.adsabs.harvard.edu/abs/2018MNRAS.475.1213S} {475, 1213}

\bibitem[\protect\citeauthoryear{{Scoccimarro}}{{Scoccimarro}}{1997}]{Scoccimarro97}
{Scoccimarro} R.,  1997, \mn@doi [\apj] {10.1086/304578}, \href
  {http://adsabs.harvard.edu/abs/1997ApJ...487....1S} {487, 1}

\bibitem[\protect\citeauthoryear{{Shimabukuro}, {Yoshiura}, {Takahashi},
  {Yokoyama}  \& {Ichiki}}{{Shimabukuro} et~al.}{2016}]{Shimabukuro15}
{Shimabukuro} H.,  {Yoshiura} S.,  {Takahashi} K.,  {Yokoyama} S.,   {Ichiki}
  K.,  2016, \mn@doi [\mnras] {10.1093/mnras/stw482}, \href
  {http://adsabs.harvard.edu/abs/2016MNRAS.458.3003S} {458, 3003}

\bibitem[\protect\citeauthoryear{{Shimabukuro}, {Yoshiura}, {Takahashi},
  {Yokoyama}  \& {Ichiki}}{{Shimabukuro} et~al.}{2017}]{Shimabukuro17}
{Shimabukuro} H.,  {Yoshiura} S.,  {Takahashi} K.,  {Yokoyama} S.,   {Ichiki}
  K.,  2017, \mn@doi [\mnras] {10.1093/mnras/stx530}, \href
  {http://adsabs.harvard.edu/abs/2017MNRAS.468.1542S} {468, 1542}

\bibitem[\protect\citeauthoryear{{Sokasian}, {Yoshida}, {Abel}, {Hernquist}  \&
  {Springel}}{{Sokasian} et~al.}{2004}]{Sokasian04}
{Sokasian} A.,  {Yoshida} N.,  {Abel} T.,  {Hernquist} L.,   {Springel} V.,
  2004, \mn@doi [\mnras] {10.1111/j.1365-2966.2004.07636.x}, \href
  {http://adsabs.harvard.edu/abs/2004MNRAS.350...47S} {350, 47}

\bibitem[\protect\citeauthoryear{{Tassev}, {Zaldarriaga}  \&
  {Eisenstein}}{{Tassev} et~al.}{2013}]{Tassev13}
{Tassev} S.,  {Zaldarriaga} M.,   {Eisenstein} D.~J.,  2013, \mn@doi [Journal
  of Cosmology and Astro-Particle Physics] {10.1088/1475-7516/2013/06/036},
  \href {https://ui.adsabs.harvard.edu/abs/2013JCAP...06..036T} {2013, 036}

\bibitem[\protect\citeauthoryear{{Zaldarriaga}, {Furlanetto}  \&
  {Hernquist}}{{Zaldarriaga} et~al.}{2004}]{ZFH04}
{Zaldarriaga} M.,  {Furlanetto} S.~R.,   {Hernquist} L.,  2004, \mn@doi [\apj]
  {10.1086/386327}, \href {http://adsabs.harvard.edu/abs/2004ApJ...608..622Z}
  {608, 622}

\makeatother
\end{thebibliography}

\appendix
\section{Effect of simulation Box size on 21cm 2PCF measurements}\label{app:2PCF_boxsize}

The size of the simulation box determines the wavelength of the largest mode,
in the matter density field of our simulation. The box size can therefore affect
the matter and 21cm correlation functions, which we measure in the simulations on large scales.

\begin{figure}
\centering\includegraphics[width=9.5 cm, angle=270]{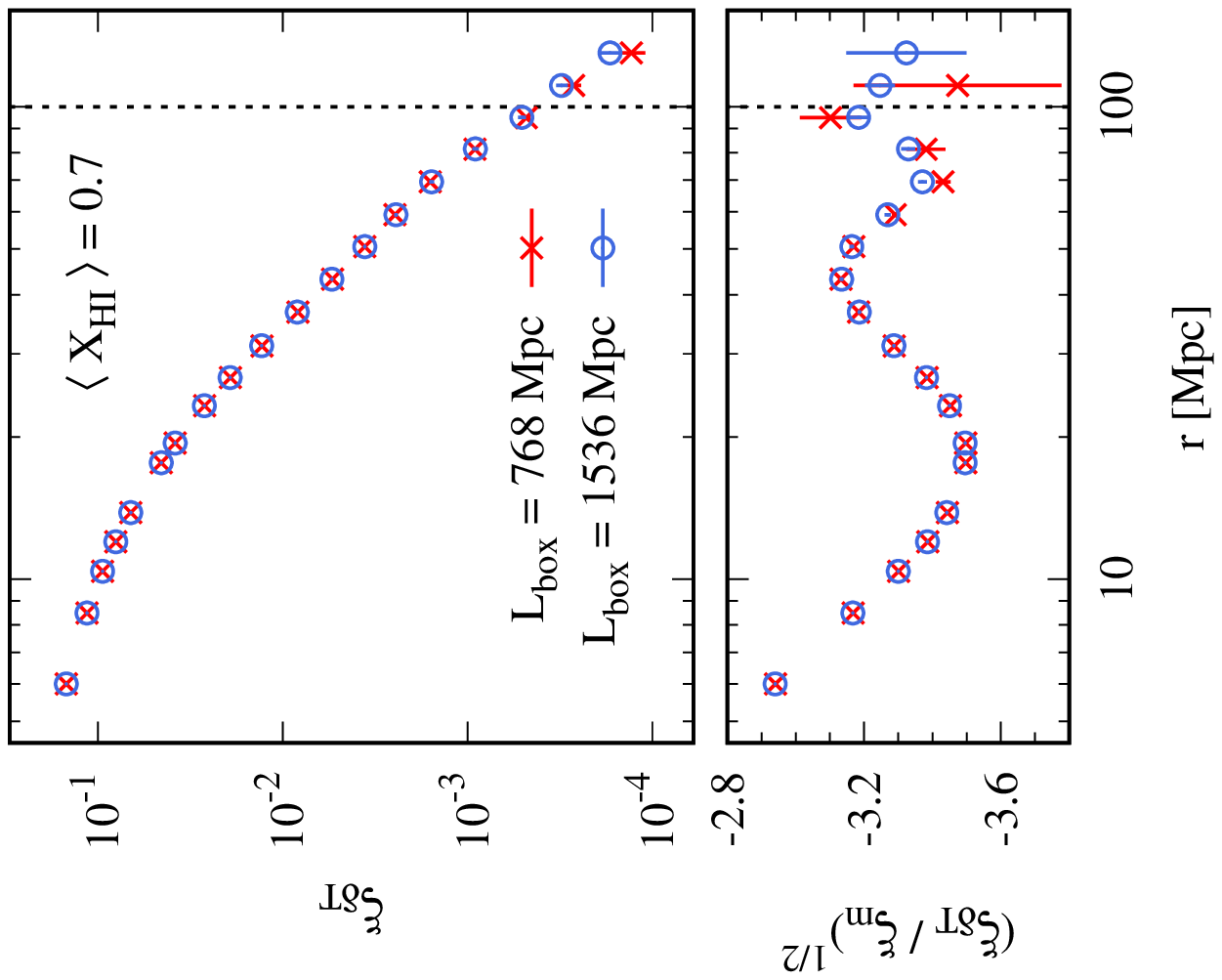}
\caption{{\it Top}: mean 21cm 2PCF, measured from $200$ ($20$) realizations
with boxes of side length $L_{box}=768$ ($1536$) Mpc at $z=9.77$ with \nf$=0.7$
(red crosses and blue circles respectively). 
{\it Bottom}: square root for the ratio of the 21cm and matter 2PCF, which
corresponds to the linear bias from equation (\ref{eq:2pc_bias}). The dotted
vertical line marks the maximum scale, used for the 2PCF bias fits.}
\label{fig:2PCF_boxsize}
\end{figure}

To test how strongly our 21cm 2PCF measurements are affected by the chosen box size
we generate $20$ realizations, each with twice the side length (i.e. $L_{\rm box}=1536$ Mpc)
than for the $200$ realizations (with the side length $L_{\rm box}=768$ Mpc) used in our analysis.
In the top panel of Fig. \ref{fig:2PCF_boxsize} we compare the 21cm 2PCF measurements
for both simulation boxes at the redshift $z=9.77$ (\nf$=0.7$). Our results show that our chosen
box size affects the 2PCF measurements on large scales $\gtrsim 100$ Mpc, while results at smaller
scales are not significantly affected.
%

\section{Relative deviations of 21cm 3PCF fits from measurements}\label{app:3pc_fit_reldiff}

\begin{figure}
\centering\includegraphics[width=11 cm, angle=270]{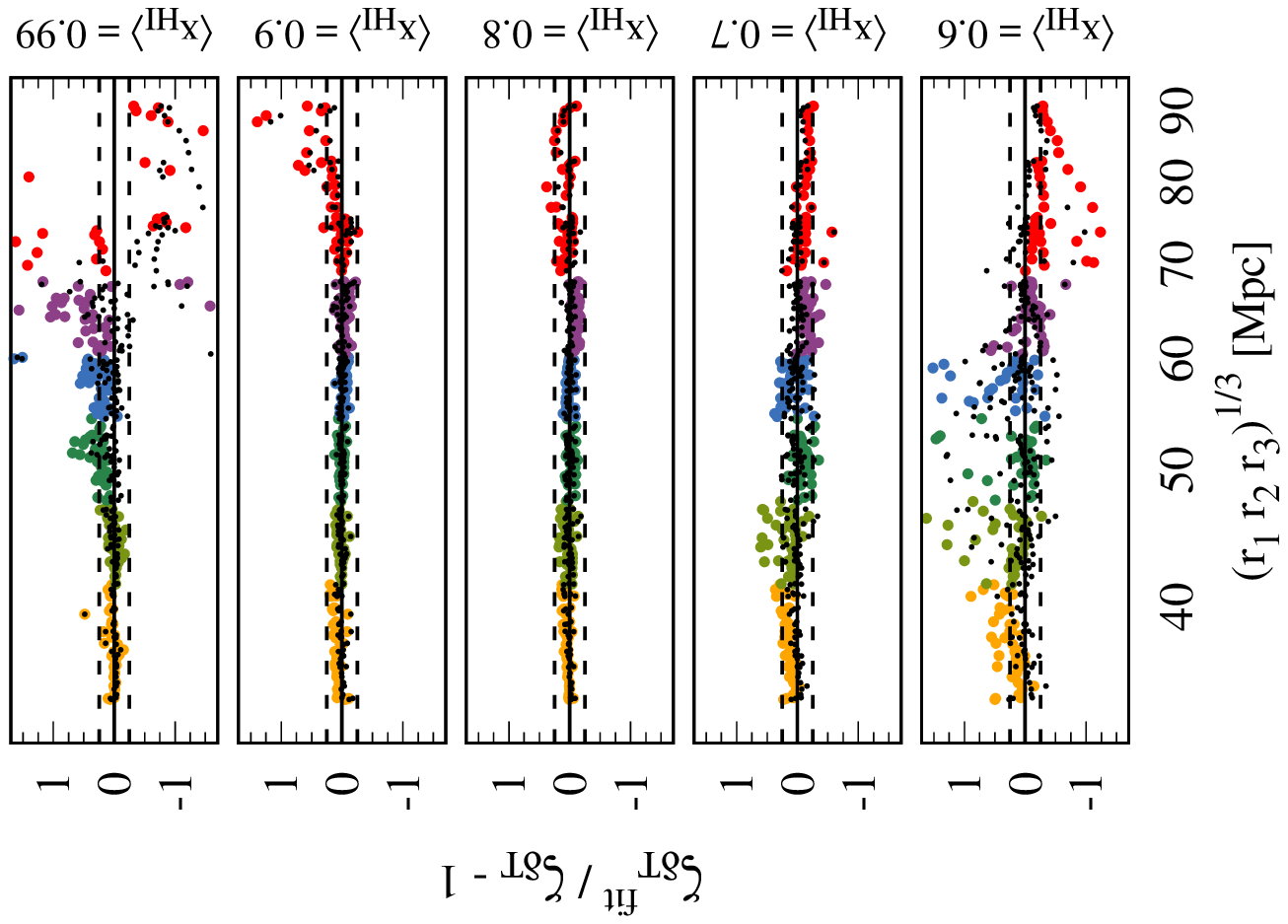}
\caption{Relative deviations between the measurements of the 21cm 3PCF from model A without RSD 
and fits to the latter based on the quadratic bias model as a function of the triangle length scale.
The panels show results for various redshifts with different global neutral fractions. The fits
are performed separately for triangles in 6 different scale bins. Each bin contains 40 triangles,
which are marked by the same colour. Black dots show results for fits which do not take into
account the off-diagonal elements of the error-covariance. Relative deviations of $\pm 25\%$ are marked
by dashed lines.}
\label{fig:3PCF_reldiff}
\end{figure}

We test the fitting performance of the bias model here by showing the relative deviations
between fits and measurements at different triangle scales and global neutral fractions
in Fig. \ref{fig:3PCF_reldiff}. This test is complementary to the test of the significance
of deviations between fits and measurements shown in Fig. \ref{fig:3pc_allconfigs}. In the
latter figure we find a better performance of the bias model at large scales. With Fig.
\ref{fig:3PCF_reldiff} we point out, that the large relative deviations can occur at small
as well as on large scales. Since the 3PCF errors increase with scales due to fewer modes
in our simulation volume, large-scale deviations are less significant.

\section{Covariances and $\chi^2$ fitting}\label{app:fitting}

\begin{figure}
\centering\includegraphics[width=5.5 cm, angle=270]{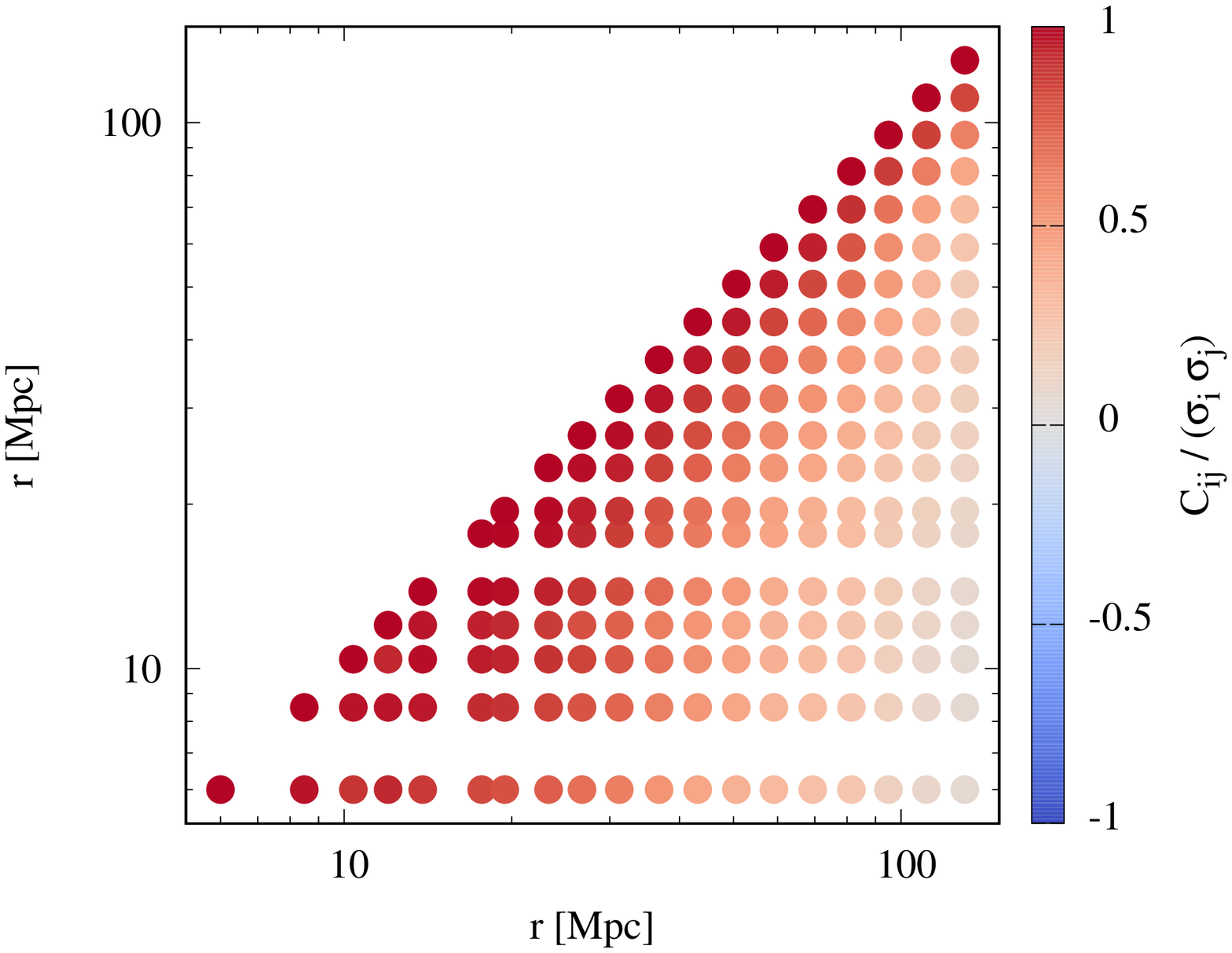}
\caption{Normalized covariance of the 21cm 2PCF for model A without RSD at \nf$=0.7$,
estimated from measurements in $200$ realizations for different scales $r$.
}
\label{fig:2PCF_cov}
\end{figure}

\begin{figure*}
\centering\includegraphics[width=14 cm, angle=270]{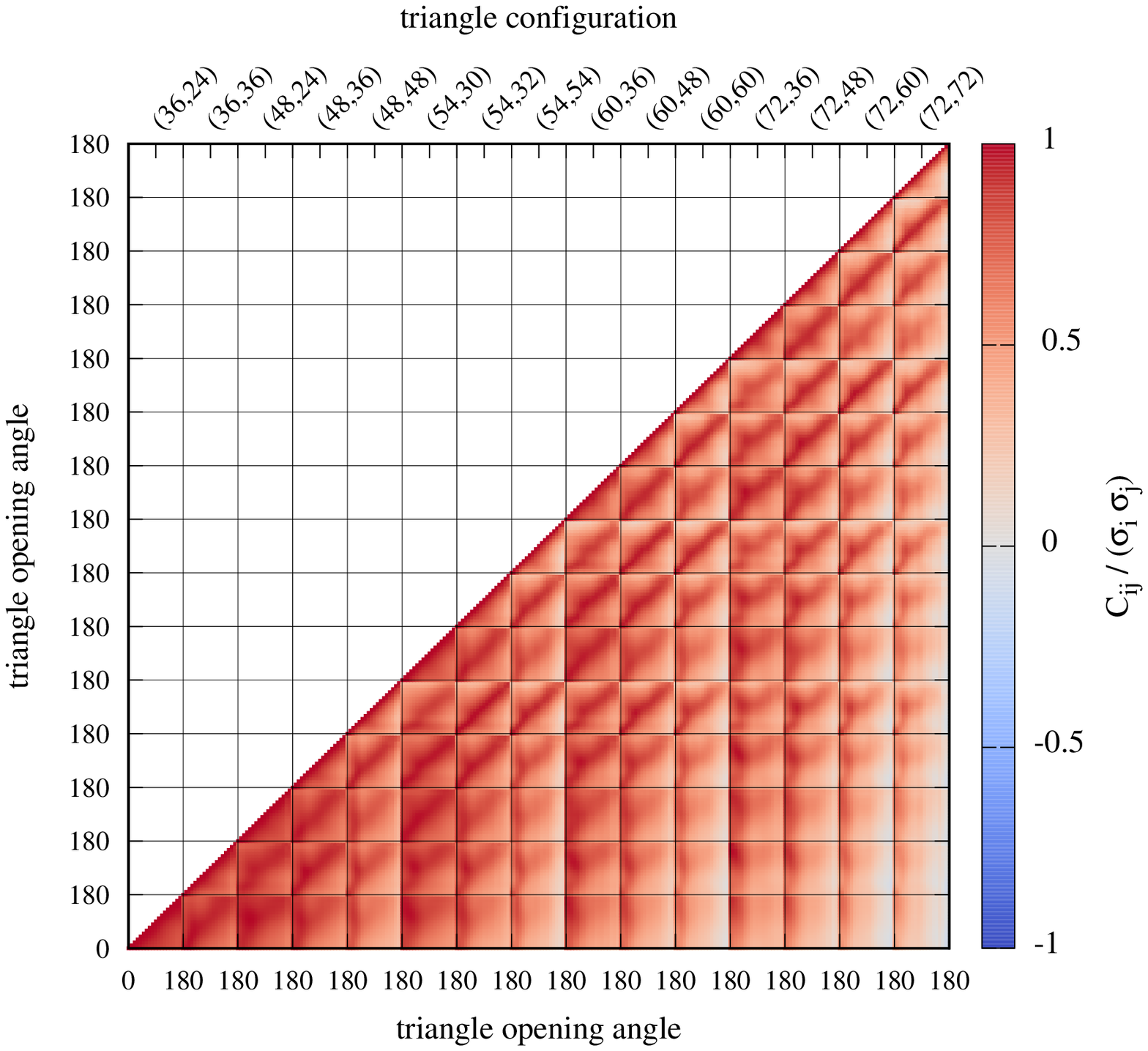}
\caption{Normalized covariance of the 21cm 3PCF for model A without RSD at \nf$=0.7$, estimated from measurements
    in $200$ realizations. Each tile shows results from different triangle
    configurations, defined by the fixed triangle legs $(r_1, r_2)$. The leg sizes are
    shown in Mpc at the top of each column.  The 3PCF was measured for each configuration at
    $18$ opening angles between $0$ and $180$ degree, leading to a total of $270$ triangles
    from all configurations.
}
\label{fig:3PCF_cov}
\end{figure*}

We fit the bias model predictions for the 21cm 2PCF and 3PCF (equation (\ref{eq:2pc_bias}) and
(\ref{eq:3pc_bias}) respectively) to the corresponding measurements by a $\chi^2$ minimization.
We therefore explore the parameter space, searching for the minimum of 
\begin{equation}
    \chi^2=\sum^N_{ij} \Delta_i {\hat C}_{ij}^{-1}\Delta_j,
\label{eq:chisq}
\end{equation}
where $\Delta_i \equiv (X^{model}_i-\langle X_i \rangle)/\sigma_i$.
The mean measurements  for a certain scale or triangle $i$
over the $N_{sim}$ realizations are denoted as $\langle X \rangle_i$, while $X_i^{model}$ are
the corresponding predictions. The variance of $\langle X\rangle_i$ is given by
$\sigma_i^2 = \langle (X_i - \langle X \rangle_i)^2 \rangle / N_{sim}$. The factor $1/N_{sim}$
accounts for the fact that we are interested in the errors on the mean measurements, rather than those
on measurements from individual realizations. The normalized covariance matrix is hence given by
\begin{equation}
	\hat C_{ij} = \langle \Delta_{i}\Delta_{j} \rangle /N_{sim},
	\label{eq:cov}
\end{equation}
with $\Delta_{i} \equiv (X_i - \langle X_i \rangle)/\sigma_i$.
%
We show examples of our $\hat C_{ij}$ measurements for the 2PCF and 3PCF in
Fig. \ref{fig:2PCF_cov} and \ref{fig:3PCF_cov}, respectively. In both cases, 
the covariance matrix has strong off-diagonal elements, in particular at small scales.
This indicates that knowledge of the covariance is essential for constraining theory models
from 21cm 2PCF and 3PCF measurements in future observations.
Based on these covariance measurements, our fits are performed for measurements from
$N<N_{sim}$ different scales or triangles, to ensure that the covariance is
invertible \citep{Hartlap07}.

\begin{figure*}
\centering\includegraphics[width=5 cm, angle=270]{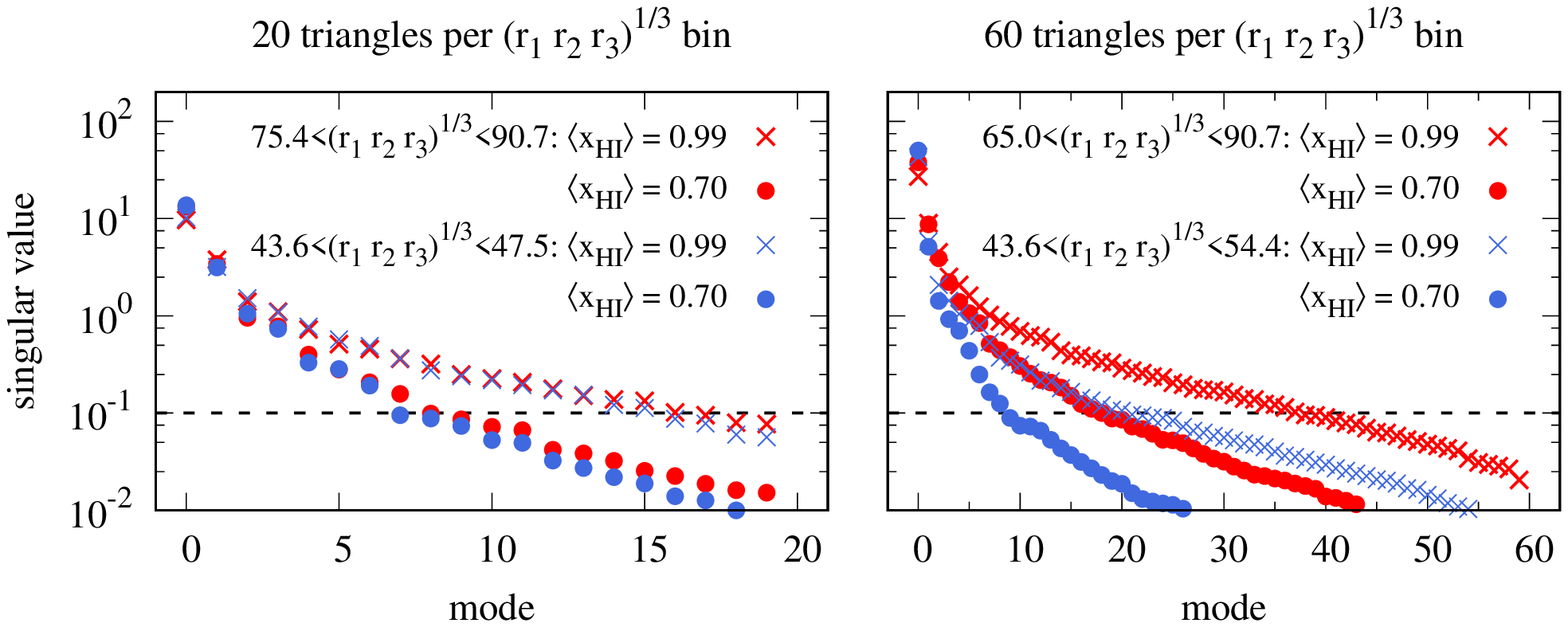}
\caption{Singular values of different modes for examples of covariance matrices used in this
work. The black dashed line marks the shot noise limit of $\sqrt{2/N_{sim}}$, below which
modes are neglected in the $\chi^2$ calculation.
}
\label{fig:3PCF_covSV}
\end{figure*}

We find that 3PCF fits based on equation (\ref{eq:chisq}) can be very sensitive
to small changes in the selected triangle sample, which may be caused by noise
in our covariance estimation from the only $200$ realizations. To reduce this noise,
we follow \citet{GazSco05} by performing a Singular Value Decomposition of the
covariance (hereafter referred to as SVD), i.e.
\begin{equation}
	\hat C_{ij} = (U_{ik})^{\dagger} D_{kl} V_{lj}.
	\label{eq:covSDV}
\end{equation}
The diagonal matrix $D_{kl} = \delta_{kl}\lambda_k^2$ consists of 
the singular values $\lambda_j$ (``SVs''), while the corresponding normalized
modes ${\bf \hat M}_i$ form the matrix $U$. The modes associated to the largest SVs
may be understood analogously to eigenvectors. The $\chi^2$ expression from
equation (\ref{eq:chisq}) can now be approximated by writing it in terms of the
most dominant modes
\begin{equation}
	\chi^2 \simeq \sum_k^{N_{mode}} ( {\bf \Delta} \cdot \hat {\mathbfit M}_k)^2 / \lambda_k^2.
	\label{eq:chisq_SVD}
\end{equation}
The elements of the vector ${\bf \Delta}$ correspond to the quantity $\Delta_{i}$,
which appears in equation (\ref{eq:chisq}). Fig. \ref{fig:3PCF_covSV} shows that $\hat C_{ij}$
is typically dominated only by a few modes.
Assuming that the modes with the lowest SVs can be associated with measurement noise,
we use only SVs with values larger than the sampling error estimate
 (i.e. $\lambda^2 \gtrsim \sqrt{2/N_{sim}}$) for our $\chi^2$ computation,
as suggested by \citet{GazSco05}. The number of selected modes is hence the
degree of freedom in our $\chi^2$ estimation, i.e. d.o.f. = $N_{mode} < N_{bin} < N_{sim}$.
Note that the description above matches in various parts of Section 2.4 in \citet{Hoffmann17},
who did a similar analysis for the halo 3PCF.

\bsp   
\label{lastpage}
\end{document}